\documentclass[10pt]{article}

\usepackage{amssymb}
\usepackage{amsmath}
\usepackage{hyperref}
\numberwithin{equation}{section}

\usepackage{graphicx}

\begin{document}

\title{ Bohmian Non-commutative Dynamics: \\{\Large History and New Developments. }}
\author{B. J. Hiley.}
\date{TPRU, Birkbeck, University of London, Malet Street, London WC1E 7HX\footnote{b.hiley@bbk.ac.uk}}

\maketitle

\begin{abstract}
  The reality of Bohm's intellectual journey is very different from what is often claimed by the proponents of ``Bohmian Mechanics" and others as we will explain in this paper.  He did not believe a mechanical explanation of quantum phenomena was possible.  Central to his thinking, and incidentally to Bohr's also, was the notion of `unbroken wholeness', a notion that is crucial for understanding quantum properties like quantum nonlocality.  His proposals were based on a primitive notion of `process' or `activity', producing a more `organic'  approach to quantum phenomena.   He published many papers outlining these new ideas, some plausible, some less so, but was not able to develop a coherent mathematical structure to support the work.  Over the last ten years much of that missing mathematics has been put in place.  This paper will report this new work, concentrating on providing a coherent overview of the whole programme.

\end{abstract}

\section{Introduction}

Over the many years that have passed since Bohm originally made his 1952 proposal \cite{bo52}, myths have grown up as to the source of his motivation to challenge the standard interpretation of the quantum formalism. In the first part of this paper the historical evolution of Bohm's ideas will be presented, quoting from the original sources so that there can be no misunderstanding of his actual position.  In the second part, an overview of the main advances that have been made in the last couple of decades.  Some of his informal ideas have now been translated into a well defined mathematical structure which has thrown new light on his overall approach.  

\subsection{The Beginning.}

 Contrary to what has been claimed, it was not ideologically driven, nor was it an attempt to challenge the predictive successes of the standard formalism. It certainly was not an attempt to return to classical determinism within a classical mechanistic philosophy.  Yet since his death there has developed an approach under the heading ``Bohmian mechanics'', which, although using {\em some} of the same formalism, has a totally different rational from that envisaged by Bohm himself.
Indeed he came into my room at Birkbeck College one day where we worked together since the early 60s carrying a pre-print of a paper, which contained a section  headed ``Bohmian Mechanics''.   He asked ``Why do they call it Bohmian mechanics?  Haven't they understood what I have been saying over all these years?" 

What was missing in the standard approach, Bohm claimed, was an adequate notion of an independent actuality \cite{db87a}.  Bohm's original proposal \cite{bo52} was intended to be merely a preliminary step, published simply to demonstrate that, contrary to common belief at that time, an account based on an independent actuality however limited was, in fact, possible.  The remainder of his work in physics was to investigate different structures that  he hoped would lead to a better understanding of quantum phenomena.

\subsection{ The Future.}

This later work was intended to supply a radically new approach that would not only provide an intuitive understanding of the underlying reality, but, hopefully also provide new insights into how  quantum phenomena could be united with general relativity.  But first we had to start with the `52 model \cite{bo52}, taking it as it stands, exploring what insights, if indeed any, it gave us.  We collected together the work we had done together on this approach in our book, ``The Undivided Universe",  [UU] \cite{bh93a}.  We had intended to write a sequel in which a much more radical approach was to be presented.  We gave a brief sketch of some of these ideas in the last chapter, which we were hoping would form the basis of the new book, but unfortunately Bohm died just as we were finalising the UU.

It was never intended that this book was to be our final word in our research together.  I had hoped this would be clear when we wrote in the introduction,
 
 \begin{quote}However, because our interpretation and the many others that have been proposed lead, at the present, to the same predictions for experimental results, there is no way experimentally to decide between them.  Arguments may be made in favour or against any of them on various basis, which include not only those that we give here, but also questions of beauty, elegance, simplicity and economy of hypotheses.  However these latter are somewhat subjective and depend not only on the particular tastes of the individual, but also socially adopted conventions, consensual opinions and many other factors which are ultimately imponderable and which may be argued many ways.
 
	There does not seem to be any valid reason at this point to decide finally what would be the accepted interpretation.  But is there a valid reason why we need to make such a decision at all?  Would it not be better to keep all options open and to consider the meaning of each of the interpretations on its own merits, as well as in comparison with others?  This implies there should be a kind of dialogue between different interpretations rather than a struggle to establish the primacy of any one of them.\end{quote}

The aim of this paper is to explain in more detail Bohm's original motivations and also to give a brief account of some of the more interesting proposals we followed. These original proposals have been considerably extended and new results have been obtained.  Not only have we extended the original '52 work to include the non-relativistic spin, but have found a method of applying the ideas to the relativistic Dirac electron.  This has focussed on Clifford algebra techniques which we found fitted Bohm's radical idea of developing an approach that assumes an underlying process philosophy based on what he called the `holomovement' \cite{bh70}.

A considerable surge occurred when two facts became clear.  (1) The original '52 approach was seen to be intimately related to the Moyal algebra \cite{bh13}.  (2) That the Moyal algebra was isomorphic with the quantum algebra proposed by von Neumann \cite{vn31}, the algebra that led von Neumann to write to Birkoff,

\begin{quote}
I would like to make a confession which may seem immoral: I do not believe absolutely in Hilbert space any more. \cite{vn35}
\end{quote}
For an enlightened account of the development of von Neumann's ideas see Redei's paper ``Why John von Neumann did not like the Hilbert Space Formalism of Quantum Mechanics (and What he Liked Instead.)"  \cite{mr96}

\section{Quantum Nonmechanics.}

\subsection{The First Analysis.}

 Let me start this section by clearing up a very basic misconception; anybody who knew Bohm and was familiar with his work would have been aware that Bohm did not believe that quantum phenomena could be explained in terms of any kind of mechanistic philosophy except in some very limited way.  Indeed, in a footnote in his classic text, ``Quantum Theory''
  \cite{bo51a} he had already written,
  
  \begin{quote} This means that the term `quantum mechanics' is very much a misnomer.  It should, perhaps, be called `quantum nonmechanics'.\end{quote} 
 
It has been argued that his position with regard to a mechanical interpretation might have changed once he had discovered the alternative interpretation outlined in his 1952 paper \cite{bo52}, but this is not the case as his book ``Causality and Chance in Modern Physics'' [CC] \cite{bo57}, written in 1957, confirms.  I begin by quoting from p. 110

 \begin{quote} At this stage, as pointed out in Section 1, the author's [sic DB] principle purpose had not been to propose a definitive new theory, but  mainly  to show, with the aid of a concrete example, that alternative interpretations of quantum theory were in fact possible.  Indeed, the theory in its original form, although completely consistent in a logical way, had many aspects which seemed quite artificial and unsatisfactory.  Nevertheless, as artificial as some of these aspects were, it did seem that the theory could serve as a useful starting-point for further developments...\end{quote}

Then Bohm goes on to suggest that a mechanistic theory may be a good starting point from which to uncover a deeper, richer structure but the philosophy of mechanism is limited in attempting to give an adequate explanation of quantum phenomena.  He then goes on in Chapter 5 to develop, and I quote,
 \begin{quote} a more general outlook which allows a more satisfactory resolution of several important problems, scientific as well as philosophical, than is possible within the frame work of a mechanistic framework.\end{quote}

During my discussion with Bohm, he continued this theme of consistently exploring new approaches, some radical, some not so radical.  It is the relationship of these developments and their relationship to the original '52 work that I will bring out in this paper.
 
 \subsection{The Basic Problem.}
 

To fill in the background for those unfamiliar with my relationship with Bohm, I would like to point out  that Bohm and I were appointed to the teaching staff in  the Physics Department at Birkbeck College, starting on the same day in 1961.  I already had a Ph.D. in cooperative phenomena but was keen to work on the foundations of quantum mechanics. 

 Our rooms were opposite each other across a narrow corridor and Bohm would pop into my room two or three times a week during term time, and whenever we were in College together in the vacations. These meetings went on until his untimely death in 1992. In fact we had our final discussion in my room on the very day he passed away.

 During our meetings, Bohm would continue from where we left off in the previous meeting, clarifying, where necessary, aspects of the previous discussions before going on to present his latest ideas, which we would discuss at length.  These meetings covered many topics, not only on physics and science in general, but also on a much wider range of subjects involving philosophy, psychology, language and even art.  I well understand that this type of discussion is not usually encouraged in physics departments, but with Bohm things were very different. In trying to develop a radical approach to quantum phenomena, the answers would not be found simply by discussing the subject in its present context.  New ideas can only enter from much wider considerations, and believe me, Bohm did not limit the subjects we discussed.

For the first ten or so years of working together, we did not discuss the causal interpretation (the content of his original 1952 paper) \cite{bo52}, or if we did, it was merely some passing remark.  I certainly did not study in detail the two papers until much later.  We were discussing more radical approaches which were motivated by our discussions with Roger Penrose, who was on the staff in the Mathematics Department at that time.  The topic of these discussions was how to extend quantum theory to include general relativity.

When we did finally get round to discussing the content of his original `52 paper,  Bohm himself would be critical of taking the causal interpretation as {\em the} definitive interpretation of the quantum formalism even in the non-relativistic domain.  He felt there were too many problems with it in its primitive form, although it contained some features that were very enticing and worthy of discussion.  On the other hand, there were some aspects some that were not so convincing as a physically intelligible interpretation. 
 Even Bell \cite{jb87a} acknowledges that ``Bohm did not like it very much''. 
 
 I  will produce some quotes from his papers over the years supporting this claim so there can be no misunderstanding as to where this outlook originated.  I was very fortunate to able to discuss these ideas with him and was delighted when he agreed to present a record of the work we did together  in our book,  UU \cite{bh93}.  There we presented a detailed account of the new ideas that had developed out of our discussions of  the earlier work over the years, presenting an approach that I call the `Bohm interpretation'.  Not only did we present what we thought were the advantages of this later approach over the original proposals \cite{bh87},  but we also outlined some of its short comings,  finally sketching, as I have already remarked, some radical ideas that we were hoping to develop in a subsequent publication.  Nowhere did we claim that the model was entirely free of troubles.  
Indeed if we had thought it was {\em the} `final' interpretation, then we would not have written the last chapter.  There we argued that we could not remain within the Cartesian order which describes an essentially local order.  We write
\begin{quote}
In the quantum domain however this order shows its inadequacy, because physical properties cannot  be attributed unambiguously to well-defined structures and processes in space time while remaining within Hilbert space. [p. 350]
\end{quote}
We argued what is needed is a new concept of order, the implicate order in which what we have learnt from the original Bohm interpretation can be seen to arise in a coherent way from this underlying structure.   Notice that the new ideas developed out of the struggle to clarify the `52 interpretation and we felt it was necessary to clarify these as best we could before developing the more radical approach in detail.

\subsection{Back to the Early Days.}

Having set the general background, let me go back in time and start with Bohm's book Quantum Theory \cite{bo51} which was published in 1951. This book written because Bohm felt the best way to understand a difficult theory was to write a book about it!  Thus the book was an attempt to explain quantum theory from Bohr's point of view [SQM].  Many thought, at the time, that it was one of the better texts on standard quantum theory.  Even Einstein thought ``It was the best that could be done to explain the theory in its present form.''  
 
 In the book, under the section heading `The need for a nonmechanical description',  Bohm writes the following, 
 \begin{quote}....the entire universe must, in a very accurate level, be regarded as a single indivisible unit in which separate parts appear as idealisations permissible only on a classical level of accuracy of the description.  This means that the view of the world as being analogous to a huge machine, the predominant view from the sixteenth to nineteenth century, is now shown to be only approximately correct.  The underlying structure of matter, however, is not mechanical \cite{bo51a}.
 \end{quote} 
 A footnote immediately below this passage contains the quote above referring to `nonmechanics'.

It might be argued that this was Bohm's position before he wrote his 1952 paper and that the `52 paper should be taken as his dramatic conversion to a mechanical viewpoint.  Not so! Bohm told me that he did not write the original paper in order to return to a mechanical model, but to show that it is possible to provide some form of independent actuality, or ontology, countering the prevailing belief that this was not possible in quantum mechanics. The mantra at that time was `there was no alternative', a position strongly supported by  von Neumann's no hidden variable theorem \cite{vn55}.  

Indeed one does not have to take my word for his position with regard to the mechanistic philosophy, one simply has to read his book ``Chance and Causality in Modern Physics" [CC] published in 1957 \cite{bo57}, five years after the original paper \cite{bo52} was published, to find a lengthy discussion of his actual position.  Let me quickly guide you through the relevant sections.

Chapter 4 of CC, entitled ``Alternative interpretations of the quantum theory'', is devoted to an appraisal of the original `52 model.  On page 110, he writes 
 \begin{quote}At this stage, as pointed out in Section 1, the author's principal purpose had not been to propose a definitive new theory, but was rather mainly to show, with the aid of a concrete example, that alternative interpretations of the quantum theory were in fact possible. Indeed, the theory in its original form, although completely consistent in a logical way, had many aspects which seemed quite artificial and unsatisfactory. Nevertheless, as artificial as some of these aspects were, it did seem that the theory could serve as a useful starting-point for further developments, which it was hoped could modify and enrich it sufficiently to remove these unsatisfactory features. 

Such developments, which have in fact occurred, at least in part, and which are still going on, will be discussed in more detail in Section 5. Meanwhile, however, a number of largely independent efforts have been made in the same general direction by Vigier \cite{jv53}, Takabayasi \cite{tt52}, Fenyes\cite{if52}, Weizel \cite{ww53} and many others. While none of the efforts cited above has been able to avoid completely some kinds of artificial or otherwise unsatisfactory features, each of them introduces new ideas that are well worth further study. It is clear, then, that even if none of the alternative interpretations of the quantum theory that have been proposed thus far has led to a new theory that could be regarded as definitive, the effort to find such theories is nevertheless becoming a subject of research on the part of more and more physicists, who are apparently no longer completely satisfied with continuing on the lines of research that are accessible within the framework of the usual interpretation.
 \end{quote}

In section 5 of CC, he goes on to list a number of what he calls `significant criticisms' of the original model.  His first remark is to point out that there is difficulty with spin and relativity, but then he continues 
 \begin{quote} Secondly, even in the domain of low energies, a serious problem confronts us when we extend the theory given in Section 4 to the treatment of more than one electron. This difficulty arises in the circumstance that, for this case, Schr\"{o}dinger's equation (and also Dirac's equation) do not describe a wave in ordinary three-dimensional space, but instead they describe a wave in an abstract 3$N$-dimensional space, when $N$ is the number of particles. While our theory can be extended formally in a logically consistent way by introducing the concept of a wave in a 3$N$-dimensional space, it is evident that this procedure is not really acceptable in a physical theory, and should at least be regarded as an artifice that one uses provisionally until one obtains a better theory in which everything is expressed once more in ordinary three-dimensional space.
 \end{quote}
He continues 
 \begin{quote}Finally, our model in which wave and particle are regarded as basically different entities, which interact in a way that is not essential to their modes of being, does not seem very plausible.  The fact that wave and particle are never found separately suggests instead that they are both different aspects of some fundamentally new kind of entity which is likely to be quite different from a simple wave or a simple particle, but which leads to these two limiting manifestations as approximations that are valid under appropriate conditions.
 
It must be emphasised, however, that these criticisms are in no way directed at the logical consistency of the model, or at its ability to explain the essential characteristics of the quantum domain. {\em Rather, they are based on broad criteria, which suggest that many features of the model are implausible and, more generally, that the interpretation proposed in section 4 does not go deep enough}. Thus, what seems most likely is that this interpretation is a rather schematic one which simplifies what is basically a very complex process by representing it in terms of the concepts of waves and particles in interaction.[My italics]
 \end{quote}

In Bohm's paper ``Hidden Variables in Quantum Theory'' \cite{bo62}, Bohm writes 
 \begin{quote} First of all, it must be admitted that the notion of the ``quantum potential'' is not an entirely satisfactory one.  For not only is the proposed form rather strange and arbitrary, but also (unlike other fields such as the electromagnetic) it has no visible source.  This argument by no means invalidates the theory as a logical self-consistent structure, but only attacks its plausibility.
 \end{quote}
 
  He then goes on to look at the many body `guidance' conditions \newline  $p_{i} = \partial S(x_{1}\dots x_{n}\dots x_{N})/\partial x_{i}$.  [Note that this expression is in configuration space.] He then immediately remarks
  \begin{quote}
   All of these notions are quite consistently logically.  Yet it must be admitted that they are difficult to understand from the physical point of view. At best they should be regarded, like the quantum potential itself, as schematic or preliminary representations of certain features of some more plausible physical ideas to be obtained later.
\end{quote}

Bohm's feelings at that time about his original model could not be clearer.  As remarked earlier, this is why Bell claims that Bohm himself did not like his own model and why Bell continues, 
\begin{quote}But like it or lump it, it is perfectly conclusive as a counter example to the idea of vagueness, subjectivity and indeterminism \cite{jb87a}.
\end{quote}

Recall why Bell was criticising the standard interpretation of quantum mechanics [SQM] in this way. Bohr had left us with the idea that you can only talk about the results of experiment and not what is going on between measurements. Bohr writes:
\begin{quote}Éin quantum mechanics, we are not dealing with an arbitrary renunciation of a more detailed analysis of atomic phenomena, but with a recognition that such an analysis is in principle excluded \cite{nb61a}.
\end{quote}
 It is this feature of SQM that both Bohm and Bell were questioning.

\subsection{A Digression to what lay ahead.}

At this point I want to amplify my explanation of why Bohm made his comment quoted at the beginning of this paper expressing surprise of the use of the phrase ``Bohmian Mechanics'' and why Bohm felt we had to go beyond mechanistic explanations.  The clearest source explaining his position is, once again ``Causality and Chance'' \cite{bo57} which contains many of the germs of the ideas that he and I later developed in new directions.  He makes it very clear towards the end of chapter 4 that the emergence of DB(52) confirmed his earlier views that quantum processes demanded we give up looking for mechanistic explanations.

Section 9, page 126 of CC has the heading ``Alternative interpretations of quantum theory and the philosophy of mechanism''.  It begins,
 \begin{quote}The consideration of the alternative interpretation of the quantum mechanics discussed in this chapter [sic DB(52)] serves to show that when one divests the theory of the irrelevant and unfounded hypothesis of the absolute and final validity of the indeterminacy principle, one is led to an important new line of development, which strikes at the heart of the entire mechanistic philosophy.\end{quote} 

In chapter five, entitled ``More general concepts of natural law'', there is a detailed criticism of the philosophy of mechanism. I cannot do justice to this chapter by attempting to summarise the arguments.  The points he makes are subtle and deep.  It draws on Bohm's experiences of plasma physics, of quantum theory, of quantum field theory, of particle physics and of relativity.  

He emphasises what for him was the most significant new feature of QM, namely, the interconnectivity of things or `wholeness'. He writes, 
 \begin{quote}First of all, we note that the universal interconnection of things has long been so evident from empirical evidence that one can no longer question it \cite{bo57b}.
 \end{quote}
 
 He is not talking about a mechanistic interconnection which is nothing more than an interaction between the fundamental entities that go to make up the system. Rather it is an interconnection that sustains the very entities themselves, and defines their properties.  Alter the background and the substructure of these interconnections and the entities themselves transform into new entities.  This is not a world of particles with well-defined properties interacting through mechanical forces.  Something much more radical is involved. 
 \begin{quote}A fundamental problem in scientific research is then to find what are the things that in a given context, and in a given set of conditions, are able to influence other things without themselves being significantly changed in their basic qualities, properties, and laws.  These are, then, the things that are, within the domain under consideration, autonomous in their essential characteristics to an adequate degree of approximation \cite{bo57c}.
 \end{quote}

One sub-section in this last chapter is entitled ``The process of becoming''.  In this section he introduces a radically new idea which permeates all his later writings.  
 \begin{quote}Thus far, we have been discussing the properties and qualities of things mainly in so far as they may be abstracted from the processes in which things are always changing their properties and qualities and becoming other things.  We shall now consider in more detail the characteristics of these processes which may be denoted by ``motion'' \cite{bo57d}.
 \end{quote}
 
 I personally did not like the word `motion' and in our subsequent discussions it became `movement', a primitive term which Bohm described by the phrase ``movement is what {\em is}".   The term ``movement" ultimately gave way to a new term, ``holomovement'' \cite{bh70}, the word being chosen to emphasise the notion of wholeness which Bohm felt to be the essential feature of quantum processes.
In this approach he is already giving up the notion that physics is about particles-in-interaction, which was the basic assumption used in DB52. Rather he is assuming the notion of {\em  process}, {\em flux}, or {\em the process of becoming}, is fundamental and underlies all physical phenomena.  
 Objects, such as particles, and even fields are to be abstracted from this underlying process.    
 
 Over the years these ideas were developed in a whole series of publications by Bohm himself \cite{db71}, \cite{db73}, \cite{db80}.  He was still writing about this interconnectivity, wholeness and its implications just before he died \cite{dbdp87}  \cite{bh93}.  

\subsection{Return to the Discussion of Bohm's Early Proposals.}

We have moved too far ahead, so let me return to the specific topic of Bohm's original 1952 proposals.  What his paper established was that it is possible to provide a formalism that could give a description of quantum phenomena in terms of an independent actuality unfolding in time which takes place without the need for any human intervention whatsoever.  

His initial proposals were very simple and emerged from an exercise he was going through using the WKB approximation.  He noticed that if you truncate a series solution for the wave function sufficiently, you obtain an expression that can be given an interpretation in terms of classical mechanics, that is of a particle following a trajectory, albeit a modified classical trajectory.  In other words in this second order approximation you can retain the concept of a particle following a well defined trajectory.  Why then when you do not truncate the series, are you forced to abandon all notions of a particle following a well defined trajectory?  At precisely what stage of the approximation are you forced to give up such a notion?

If you take the WKB polar decomposition of the wave function, $\psi(x,t)=R(x,t)\exp[iS(x,t)/\hbar]$, and put it in the Schr\"{o}dinger equation without any approximations, you get a complex equation which can be separated into its real and imaginary parts.  The real part reads
\begin{eqnarray}
\partial_tS(x,t) +(\nabla S(x,t))^2/2m +Q(x,t) +V(x,t)=0	\label{eq:sch}
\end{eqnarray}
where $Q(x,t)=\hbar^2\nabla^2 R(x,t)/2mR(x,t)$. To those who are familiar with the classical mechanics will recognise immediately that if $Q=0$ then the equation becomes the well known Hamilton-Jacobi equation, provided the phase $S(x,t)$ is replaced by the classical action. Please note that contrary to some discussions, this term, $Q$, is {\em not added}.  It appears directly from the Schr\"{o}dinger equation itself. 

If we regard equation (\ref{eq:sch}) as a modified Hamilton-Jacobi equation and assume the canonical relations $p=\nabla S$ and $E=-\partial_tS$ still hold, we find the equation is simply an equation for the conservation of energy,
\begin{eqnarray}
E(x,t)=P(x,t)^2/2m + Q(x,t) + V(x,t).		\label{eq:conenergy}
\end{eqnarray}
Thus we see that $Q(x,t)$ is some new quality of energy that only appears in the quantum domain and its appearance accounts for the difference between the classical and quantum behaviour of a particle.  

I have always been surprised at the reluctance to take the presence of this term seriously.  Since it comes directly from the real part of the Schr\"{o}dinger equation, with nothing added it must be taken seriously.  Why does it appear? Agreed at first sight it seems a rather strange object, but that is no reason to dismiss it.  Surely we should try to explore its meaning by examining its structure and how it behaves in various experimental situations.   In other words, a  key question at this stage should be to explore the physical meaning of this new form of energy. 

One objection lies in calling it the `quantum potential', implying it is some new form of classical-type potential.  It has very different properties from a classical potential.  It has no external source and it need not fall off with distance. It is, in a sense, an `internal energy'.  On the other hand, it contains information about the experimental conditions, which is good as it fits exactly what Bohr wrote:-
\begin{quote}
I advocate the application of the word phenomenon exclusively
 to refer to the observations obtained under specific circumstances,
 including an account of the whole experimental arrangement \cite{nb61b}.
\end{quote}
It was this feature that led us to propose that it was an `information potential', again using the notion of `forming from within'.  Thus it might not be a potential that produces a `force', but it is a potentiality for giving rise to a form of behaviour without the need for a concept of `force'.  All of these ideas emerged from detailed calculations carried out in the 70s.

  Indeed in the fifties when Bohm was originally discussing the quantum potential, it  was not possible to carry out this exploration because it requires considerable  numerical work to perform the necessary calculations, so the physical origins of this quantum energy could not be properly explored. As a consequence the lack of physical motivation and the fact that we must use configuration space, was sufficient reason for Bohm to temporally abandon BM(52).  Indeed, as I have already remarked, from the early sixties until the mid-seventies Bohm did not even mention BM(52) in my company. We were discussing more radical approaches based on the notion of process discussed above \cite{dbsb},\cite{dbkyo}

It was two of our then research students, Chris Philippidis and Chris Dewdney, who, in the mid-seventies asked me why we never discussed BM(52).  It may be hard to understand but this was the first time I got round to actually reading BM(52) in detail, even though I had been working with Bohm since 1962! The reason was very clear.  As I have already remarked we had Roger Penrose in the Maths Department who was discussing his new ideas on twistor theory and I was much more interested in Clifford algebras at that stage.  The questions raised by the two Chris's drew my attention to the `52 work and we started a detailed reappraisal and further exploration of BM(52).  

Let me make it clear that BM(52) had all the mathematics in place to calculate trajectories and show how quantum `interference' effects can be explained by particle following trajectories, but detailed calculations had not been carried out.  Our group at Birkbeck were the first to calculate and examine in detail the trajectories for many different experimental situations. We learned later that similar calculations had been carried out  by Hirschfelder, Christoph and Palke \cite{jhacwp} and by Hieschfelder, Goebel and Bruch \cite{jhcglb}. 
We had also calculated expressions for the quantum potential so that we could see exactly how it worked to produce the trajectories to be consistent with standard quantum mechanics.  For the first time we began to see what its properties were and how it worked.  This work gave many interesting insights to quantum phenomena that was not possible with the standard formalism.

Much of the hard work was done by our research students who included Chris Dewdney, Chris Philippidis, Peter Holland, Fabio Frescura, Pan Kaloyerou.  Later I was joined by Melvin Brown, Owen Maroney, David Robson and Robert Callaghan and others who have continued to carry the work forward, by exploring different aspects of the approach.

 We started by keeping things simple.  We assumed there was a local `particle' using the Bohm formalism to calculate possible trajectories that a particle follow in order to produce the bunching needed to account for the experimental `interference' patterns.  Our first set of results appeared in \cite{pdh}, \cite{cdbh}.  
  
  Initially we explored these trajectories, as well as the corresponding quantum potentials, without any preconceived ideas as to what it might all mean for the physics. We looked at many different experimental situations in order to obtain a comprehensive picture of what was going on. It was only after we had done all the calculations that we struggled with the meaning of quantum non-locality \cite{bhnonl}, with the delayed choice experiment \cite{dbcdbh}, with the Zeno paradox \cite{bh93}.  We got `inside' the approach; it became part of us.  Yes it worked and we found out exactly {\em how} it worked, but at the end of the day we are still faced with the question ``What does it all mean?''  Can we keep the the simple classical view of physics going, a view that is so badly wanted by some?  Our conclusion was that you could {\em not} keep it simple.  It just did not hang together.

Our numerical work and discussions re-ignited Bohm's interest in his earlier work.  In 1975 we published what I thought was a turning point in our attempts to understand what could possibly be going on \cite{bhnonl}.  I think it was one of the most important papers on the subject we published together on this subject.  It was called ``On the intuitive understanding of nonlocality as implied by quantum theory''.  We had over 400 requests for reprints and it was reprinted in a collection of essays {\em Quantum Mechanics, a Half Century Later} \cite{jlmp}.  In that paper we developed interconnectedness into `the wholeness of form' which eventually led on to the concept of `active information',  a phrase first used in Bohm and Hiley \cite{dbbh82}. 
  
The development of these ideas carried us from our inherited mechanistic philosophy to a radically new philosophy based on process and wholeness. This more `organic' view didn't come easily to us, well certainly not to me.  We argued back and forth about these ideas for years!  I cannot convey the agonies we went through in all those discussions over the years.  We didn't just confine our attention to the physics handed down to us by our contemporaries.  We discussed its implications, not only all the fundamental branches of physics, but mathematics, philosophy (analytic and continental, ancient and modern); we discussed biology, linguistics and psychology.

 Our ideas were not developed in the sixties while sitting lotus position, chilled out in some exotic location! They were teased out mainly in the heart of London! We discussed with our colleagues and visitors from overseas. We even persuaded a colleague in experimental physics to make some experiments on nonlocality \cite{jldb}, and this was before Aspect \cite{aa}! As these discussions developed Bohm believed that with the introduction of the notion of `active information', we at last had a physical reason for using configuration space and, in consequence an intelligible physical explanation for the quantum potential.  But more than that, we believed that by the eighties we had the potential foundations for going beyond a mechanistic view of the quantum formalism.

We summarised all these discussions in our paper ``An ontological basis for the quantum theory'' \cite{bh87}.  In the same issue we showed how the same ideas could be extended to bosonic fields. In this we were helped by Pan Kaloyerou who made a significant contribution to the work.  Bohm himself wrote a paper ``A Realist View of Quantum theory'' \cite{db88}.   This paper repeats much of what appeared in the previous paper, placing the quantum potential central to the physical interpretation. Bohm believed it was the notion of active information that took us beyond BM(52) and onto a physically meaningful underpinning of configuration space and hence of the quantum potential.  

In the many-body system, the non-locality in the quantum potential revealed a striking new feature in quantum processes.  Bohm writes, 
  \begin{quote}
  While nonlocality as described above is an important new feature of the quantum theory, there is yet another new feature that implies an even more radical departure from the classical ontology, to which little attention has been paid thus far.  This is that the quantum potential Q depends on the {\em quantum state} of the whole system in a way that cannot be defined simply as a pre-assigned interaction between all the particles. 
  \end{quote}
   He then goes on to say, 
   \begin{quote}
   But in the causal interpretation of quantum mechanics, this {\em interaction} [the quantum potential] depends upon the wave function of the entire system, which is not only contingent on the state of the whole but also evolves in time according to the Schr\"{o}dinger equation.  Something with this sort of independent dynamical significance that refers to the whole system and that is not reducible to a property of the parts and their inter-relationships is thus playing a key role in the theory.  As we have stated above, this is the most fundamental new ontological feature implied by quantum theory.[Bohm's italics]
   \end{quote}
 
All of these ideas were collected together and re-presented in a broader context in our book \cite{bh93}.  It is the first eight chapters that essentially defines what I have called the Bohm interpretation BM(BH) and should be considered Bohm's last words on the subject.  I say `last words' because we just had completed the final draft and were just waiting for an opportunity to take a last look at one or two chapters when Bohm died.  I decided at that stage to leave the manuscript in the form that we had last discussed it and merely remove any typos.  We were planning to write up a follow up with a more detailed account of his more radical ideas, a sketch of which formed the content of the last chapter.  Unfortunately we never got round to writing anything down and I have since been pulling together the ideas and extending some of the underlying mathematics.  For the remainder of this paper I would like to explain this background and to briefly summarise some of the latest developments.

\subsection{Bohm's More Radical Ideas.}

As I remarked in section 2.2, when I first started working with Bohm we did not discuss his `52 work for the first ten years.  He was much more interested in exploring a more radical approach in which {\em process} was taken as basic.  I am here referring to the quotation above which refers to the notion of the `holomovement'.  The basic assumption was that quantum processes could not adequately be described by particles/fields interacting in space-time. Rather there is a deeper underlying process from which, not only do the particles and fields emerge, but this process is the source of the structure of space-time itself. 

In an important paper entitled ``Time, the Implicate order and Pre-space" \cite{db86} Bohm wrote
\begin{quote}
My attitude is that the mathematics of the quantum theory deals {\em primarily} with the structure of the implicate pre-space and with an explicate order of space and time emerges from it, rather than with movements of physical entities, such as particles and fields. (This is a kind of extension of what is done in general relativity, which deals primarily with geometry and only secondarily  with the entities that are described within this geometry.)
\end{quote}

Rather than continuing the tradition of differential geometry, it seemed that for quantum phenomena,  geometric algebras were a more appropriate tool.  Indeed in the same article Bohm continues,
\begin{quote}
The fundamental laws of the current quantum-mechanical field theory can be expressed in terms of mathematical  structures called \newline `algebras' (indeed, only three kinds of algebras are needed for this purpose, Bosonic, Clifford algebras, and Fermionic algebras).
\end{quote}
He then goes on to sketch  how these  structures could provide the formal mathematics with which a more rigorous scientific approach to the philosophic ideas of the implicate-explicate order could be developed.

Unfortunately at that stage, as I now begin to see, we had not fully understood the nature of some of the mathematics that was needed.  Since then I have had the good fortune to meet a number of first class mathematicians who have pointed me in the right direction to remedy this.  In particular I have to thank Maurice de Gosson who has helped me to understand the important role symplectic geometry plays in physical phenomena, particularly in the interphase between interference phenomena and ray dynamics, essentially a rigorous mathematical treatment of what is loosely called `wave-particle duality'.  In algebraic form this leads to the    symplectic Clifford algebras \cite{ac90} and without this structure, the orthogonal Clifford algebra that we had been exploring provides only half the structure we need.  

Bohm's intuition in the quote immediately above was partially right since each of these algebras can be constructed by using (a) pairs of fermionic or Grassmann algebras (these underpinning the orthogonal Clifford) and (b) pairs of bosonic algebras (which underpin the symplectic Clifford).  These Clifford algebras are particular examples of von Neumann algebras \cite{vj03}, which lie at the heart of what is called {\em algebraic quantum field theory} \cite{rh92}.  The structure of this theory has been developed considerably since the days when we started on our exploration and, in consequence, we now have a much richer set of formal mathematical tools at our disposal.  While these provide new detailed techniques to continue our exploration, they do not negate the overarching  philosophical ideas -- in fact they add to the general coherence of the whole approach.

These general ideas of ours go back to the late sixties and early seventies when Bohm, Penrose, Kronheimer, Geroch and myself, together with a few others, were discussing how we would incorporate general relativity into quantum theory.  What stimulated Bohm and myself was the spin network ideas that Penrose was developing at the time \cite{rp71}.  This eventually led Penrose into his twistor programme \cite{rp67}.  It turned out that the twistor is, in fact, the semi-spinor of the conformal Clifford algebra and it was one of the   facts  that started me exploring orthogonal Clifford algebras in the first place.  The conformal Clifford contains the Dirac Clifford as a sub-algebra, an algebra independently discovered by Dirac as the relativistic version of the Schr\"{o}dinger equation that naturally includes the spin of the electron-- the famous Dirac equation.  This, then,  was my original motivation to start an investigation of the formal properties of the orthogonal Clifford algebras. 

Although I must confess that I was far more interested in the formal side of the mathematics than was Bohm, he nevertheless provided considerable insight as to how these algebraic structures provide a deeper understanding of quantum phenomena.  Before I had got to grips with the mathematical details, 
Bohm's own publications discussing the algebraic approach began in the sixties when I joined him at Birkbeck College. His first paper on this approach appeared in 1965, entitled ``Space, Time and the Quantum Theory Understood in Terms of a Discrete Structure Process'',  a paper that was very influential in the subsequent development of our ideas \cite{dbkyo}.  That was followed by ``Space-time Geometry as an abstraction from Spinor Ordering'' in 1971 \cite{db71}, linking in the Clifford algebras for the first time. 

 It was at this time that we were joined by Fabio Frescura from South Africa who wrote a PhD thesis ``On the use of Boson and Fermion Algebras in Quantum Mechanics" that we set about applying these algebras to Bohm's implicate order \cite{ffbh80} \cite{ffbh80a}.  Bohm and myself then published
  ``Generalization of the Twistor to Clifford Algebras as a basis for Geometry'' in 1984 \cite{dbbh84}.  All of these were technical papers discussing a notion of what we called `pre-space', the implicate order providing the essential background to what we were doing.  In the same volume Frescura and I published the details of how to generate the orthogonal and symplectic Clifford algebras from pairs of boson and fermion algebras \cite{ffbh84}.

\section{New Results: Spin and the Dirac Particle.}

\subsection{Origins of the Bohm Momentum and Bohm Energy.}

The question that naturally arises here is ``What has all of this got to do with the original Bohm model?" Surely the whole rational behind the Bohm model was to re-establish the role of the `particle' evolving in space-time.  Of course, this is but one way to attempt to describe an actual physical process.  In fact that was exactly what Bohm himself proposed in his original paper.  However, as we have seen, he soon realised that this approach was rather simplistic and raised too many difficulties.   The notion of a localised point-like particle carrying a set of pre-established localised properties  is very much a classical notion and does not fit comfortably with quantum theory in general and quantum nonlocality in particular.

In our paper entitled ``On the Intuitive Understanding of Nonlocality as Implied by Quantum Theory"  \cite{bhnonl} we write:-
\begin{quote}
Rather, the basic qualities and relationships of all the ``elements" appearing in the theory are now seen to be generally dependent on the state of the whole, even when these are separated by distances of macroscopic order. 
\end{quote}

The notion of pre-established local properties was one of the key assumptions recently used by Leggett \cite{al03} to derive an identity that a theory must satisfy in order to be called a {\em nonlocal realist theory}.  This identity was shown by Gr\"{o}bleacher {\em et al} \cite{sgma07} to lead to what they called a generalised Leggett-type inequality.  This was then tested experimentally and it was shown that quantum systems violated this inequality. 

 However  they noted in passing that, in the case of the Bohm model, neither of the two separating entangled spin-half particles initially carried any individual spin components in the entangled state when the total angular momentum of the combined state was zero. As was beautifully demonstrated in the paper by Dewdney {\em et al} \cite{cdjv88}, the spin component were shown to develop during the measurement process on one of the spatially separated particles, highlighting the non-local feature of quantum theory.
 Gr\"{o}bleacher {\em et al}  therefore concluded that their experiment did not apply to the Bohm model as the Leggett criteria insisted that the particle carried all its intrinsic properties with it.  In other words a spin-half particle  possesses its spin in all circumstances. 

 Thus to emphasise this important point again, this model makes clear that the spin of the individual particles in an entangled pair was `created' by the quantum torque as one or other of the particles passed through a Stern-Gerlach magnet, a process that left the particles in a un-entangled state. This torque ensures conservation of angular momentum.  Thus it is the measurement that changes the whole process in a global way illustrating what we called the participatory nature of a quantum measurement. It is this notion of participation that forms a key ingredient of quantum processes, a feature to which we drew attention on p. 6 of our book, `The Undivided Universe' \cite{bh93}.  The participatory nature of the measurement was a view  shared by Wheeler \cite{jw91}

This leaves us with the question as to how we can discuss the behaviour of individual particles, which seem to be the basic form used in the Bohm model.  We answer this by writing:-
\begin{quote}
However, when the wave function can be expressed approximately as a product of functions of coordinates of different ``elements", then these latter will behave relatively independently. But such a relative independence of function is only a special case of general and inseparable dependence. So we have reversed the usual classical notion that the independent ``elementary parts" of the world are the fundamental reality and that the various systems are merely particular contingent forms and arrangements of these parts. \cite{bhnonl}
\end{quote}

We are not alone in noticing this point.    Weyl in his classic book \cite{hw28} made a similar point when discussing the technicalities involved in entangled states  of many electron atoms.  Starting with product states,  he writes
\begin{quote}Ê
The reduction of the product representation, ${\cal R}^f$, into its anti-symmetric, $\{{\cal R}^f\}$, and symmetric, $[{\cal R}^f]$, parts involves relationships which frustrate any attempt at description in terms of our old intuitive pictures with their orbits and billiard-ball electrons. ÊBut the difficulty enters already with the general composition rule, according to which the manifold of possible states of a system composed of two parts is much greater than the manifold of combinations in which each of the particle systems is itself in a pure state.
\end{quote}

Why then does the  simple Bohm model actually work?  The answer to this question will emerge later in this paper, but let us start by recalling that in the original paper \cite{bo52}  Bohm  arbitrarily replaced the classical action, $S$, used in the canonical relation $P_B=\nabla S$, by the phase of the wave function, driving at what is generally called the `guidance' condition.  I call this momentum the Bohm momentum.  What is the justification for identifying the phase with the classical action?

Bohm also used the canonical relation $E_B=-\partial_t S$, giving what I call the Bohm energy.   Bohm, himself,  regarded these relations to be `subsidiary conditions' \cite{db53}, the notion of `guidance' being irrelevant in the algebraic approach as we will see.

 In thinking about how this approach may be related to the algebraic quantum field approach, I decided to look at the energy-momentum tensor constructed from the Schr\"{o}dinger field.  I found that the fourth component of this tensor, $T^{\mu 0}$, immediately gave me the exact expressions for the Bohm energy and the Bohm momentum.  No appeal to any classical theory was needed to establish this result and thus there is no need to identify the classical action with the phase of the wave function. We now have an explanation of the physical origins of  $P_B$ and $E_B$ coming from quantum field theory.  Equation (\ref{eq:conenergy}) then shows that the quantum potential is necessary to conserve energy since we have only used part of the energy-momentum tensor.
 
 Here we will assume, initially, that the energy of the particle is below the pair creation threshold.  If we remove this restriction, our theory becomes a many-particle theory and we are able to handle particle creation and annihilation. This step immediately opens the way to answering the criticism that the Bohm model cannot handle pair creation.  It can, but an explanation of this extension will take us too far off the purposes of this paper so we will not pursue this idea further here.  
 
 Since we no longer have to derive $P_B$ and $E_B$ by identifying the classical action with the phase of the wave, we have been able to extend the Bohm approach to Pauli and Dirac particles.  In fact by using the Pauli and Dirac fields  in their respective energy-momentum tensors, we immediately find the corresponding expressions for these variables for the Pauli and Dirac particles \cite{bhbc12}.
 
 Although Bohm himself showed how to extend his original model to the non-relativistic Pauli equation \cite{dbst55}, he was unable to extend the method he was using to the Dirac particle.  We  made an attempt to include the Dirac particle in our book \cite{bh93}, but this attempt was not completely satisfactory since it was not possible to find an expression for the quantum potential using these ideas.  The complete approach shows the nature of these shortcomings as was discussed in \cite{bhbc10a}
 
 The way the method was originally extended to the Pauli equation was to express the two-component spinor in terms of the Euler angles \cite{dbst55}.  It was then a fortunate accident that the azimuthal angle could be identified as a common phase, $S$, which was used to replace the classical action in the canonical relations.  In this case we therefore have to use {\em two} arbitrary features, firstly, replace the classical action by the phase and secondly identify the azimuthal angle as the common phase, neither of which can be justified.
 
 This procedure  does not work for the Dirac spinor with its four complex components.  Using the energy-momentum tensor removes these ambiguities simply because we no longer need to convert  classical canonical relations into quantum relations.  These relations emerge directly from the quantum formalism itself.
 
 \subsection{Algebraic Spin.}
 
 In the previous subsection we have still retained the wave function.  Now we must become more radical and turn to a fully algebraic approach.  To do this we need to use the orthogonal Clifford algebra to generalise the Bohm approach to include spin.  What we find is that the information that is normally contained in the wave function is encoded in the algebra itself, in an element of a minimal left ideal, $\Psi_L(x,t)$ \cite{ffbh80a}.  It is this move that enables the Bohm approach to be generalised so that it can be applied to include the non-relativistic spin of the Pauli particle and its relativistic generalisation, the Dirac particle.
 
 Rather than working with the individual element, we find it more convenient to work with what we have called  a Clifford density element defined by
 \begin{eqnarray}
 \rho_\psi(x.t)=\Psi_L(x,t) \Psi_R(x,t)		\label{eq:CDE}
 \end{eqnarray}
 where  $\Psi_L(x,t)$  is a suitably chosen element of a left ideal, while $\Psi_R$  is an element of a corresponding right ideal.  These symbols replace the usual bra-ket symbols of the standard approach so that $\rho_\psi$ essentially replaces the usual density matrix.  

 The introduction of this element may appear very formal.  However there is an intuition behind it in the context of a process description.  Representing the element by a matrix, as we can, produces an array of sub-elements as was first presented and discussed in Born and Jordan \cite{mbpj25} in their `matrix mechanics' approach.  In such an array, the diagonal elements represent `being', while the off diagonal elements represent `becoming', so that we have an expression of both being and becoming in a given context.  We can regard `being' as a process that continually transforms into itself as an idempotent, $P.P=P$.
 
 It is interesting to note that the singular value decomposition theorem tells us that we can write any matrix in the form
 \begin{eqnarray}
 M=U\Lambda V^*		\label{eq:cde}
 \end{eqnarray}
 where $V^*$ is the conjugate transpose of $V$ and $\Lambda$ is a diagonal matrix.  Here $U$ and $V$ are regarded as left (right)-singular vectors of $M$. Furthermore by considering the elements to be real, $U$ and $V$ can be taken as rotation matrices.  This is the actual situation in equation (\ref{eq:CDE}).  To show this let us write 
 \begin{eqnarray*}
 \Psi_L=\psi_L\epsilon\quad\mbox{and}\quad \Psi_R=\epsilon \psi_R,
 \end{eqnarray*}
 where $\epsilon$ is the idempotent.  Thus
 \begin{eqnarray*}
  \rho_\psi(x.t)=\psi_L(x,t)\epsilon \psi_R(x,t)
 \end{eqnarray*}
   Furthermore, as Hestenes  and Gurtler  \cite{dhrg71} point out, $\psi_L$ produces a rotation, with $\psi_R$ producing a conjugate rotation.

 If we are interested only in pure states, then we choose our ideal to be minimal and write $\Phi_R=\widehat\Psi_L$ where $\widehat\Psi_L$ is  the anti-automorphism of $\Psi_L$, called Clifford reversion.  When we do this, we find $\rho_\psi^2=\rho_\psi$, the idempotent condition that signifies a pure state. 
 
 Notice in our algebraic approach, we do not have to use specific matrix representations for our algebraic elements, although that option is always open to us so that we can check how our work relates to the standard approach to spin.  In this way our results are quite general, being what we can call `representation-free', thus avoiding the confusion as to which representation to use in a given situation.

Our generalisation should not be dismissed as simply replacing the standard approach with a more mathematically difficult structure.  It is only when we use this algebra that we can find the quantum Hamilton-Jacobi equation for the Pauli and Dirac particles.  In other words, we are able to find an expression for the quantum potential in all cases, thus we have a way of generalising the Bohm approach to all situations including the relativistic case. 

This  now answers another criticism of the Bohm interpretation, namely, that it cannot be extended into the relativistic domain.  Now we will see it can.  Another way of extending the approach into the relativistic domain has already been presented in Bohm and Hiley \cite{dbbh91}, however that method did not include spin.    

The new approach gives us a way to explore how quantum non-locality can exist in ``peaceful coexistence with relativity".  This requires extending our algebraic approach to the two-particle case, the details of which will be presented in another publication.
  For now we must go on to discuss the time development of the single particle Clifford density element.

\subsection{Time Evolution in a Non-commutative Structure.}

In this subsection it is important to realise that we are dealing with a non-commutative structure, so any derivates used in the theory must be capable of acting from {\em both} sides of any element.  In other words we have a bi-module structure.  This means we must introduce {\em two} derivatives, $\overrightarrow\Delta_{(x,t)} \Psi_L$ and  $\Psi_R\overleftarrow\Delta_{(x,t)}$.   $\Delta_{(x,t)}$ is a derivative with respect to $x$ and/or $t$.  In the case of time, $t$, we simply write $\Delta_{(t)}=\partial_t$.  The time development equation is always generated by the Hamiltonian which involves $\nabla_x^2$.  The derivatives for the orthogonal Clifford cases are chosen from one of these generalised Dirac derivatives
\begin{eqnarray*}
\nabla&=\frac{e}{2}\sum_{i=1}^{3}\partial_{x_{i}}\hspace{1cm}\mbox{Schr\"{o}dinger}\\
\nabla&=\sum_{i=1}^{3}\sigma_{i}\partial_{x_{i}}\hspace{1cm}\mbox{Pauli}
\hspace{1cm}\\
\nabla&=\sum_{\mu =1}^{3}\gamma_{\mu}\partial x^{\mu}\hspace{1cm}\mbox{Dirac}\hspace{1cm}
\end{eqnarray*}
Thus we have both a left and a right Hamiltonian  given by 
\begin{eqnarray*}\overrightarrow{H}=H(\overrightarrow{\nabla}, V, m)\quad\mbox{and}\quad \overleftarrow{H}=H(\overleftarrow{\nabla}, V, m).
\end{eqnarray*}
This means that we must construct our time evolution equations from the four derivatives 
\begin{eqnarray*}
(\partial_t\Psi)\Psi_R;\quad\Psi_L(\partial_t\Psi_R);\quad({\overrightarrow \nabla}\Psi_{L})\Psi_{R};\quad \Psi_{L}(\Psi_{R}{\overleftarrow \nabla}).
\end{eqnarray*}
Rather than treat these two derivatives separately we will form two equations 
\begin{eqnarray}
i[(\partial_{t}\Psi_{L})\Psi_{R}+\Psi_{L}(\partial_{t}\Psi_{R})]=i\partial_{t}\hat{\rho}=(\overrightarrow{H}\Psi_{L})\Psi_{R}-\Psi_{L}(\Psi_{R}\overleftarrow{H})
				\label{eq:conprob}
\end{eqnarray}
and
\begin{eqnarray*}
i[(\partial_{t}\Psi_{L})\Psi_{R}-\Psi_{L}(\partial_{t}\Psi_{R})]=(\overrightarrow{H}\Psi_{L})\Psi_{R}+\Psi_{L}(\Psi_{R}\overleftarrow{H})
					\label{eq:anticom}
\end{eqnarray*}
The first of these equations \eqref{eq:conprob} can be written in the more suggestive form
\begin{eqnarray}
i\partial_{t}{\rho_\psi}=[H,\rho_\psi]_{-}
				\label{eq:conprob2}
\end{eqnarray}
This equation has the form of Liouville's equation and can be shown to correspond exactly to the conservation of probability equation.  A similar derivation has been given by Hirschfelder \cite{jh78}

Equation \eqref{eq:anticom} can be written in the form
\begin{eqnarray}
i[(\partial_{t}\Psi_{L})\Psi_{R}-\Psi_{L}(\partial_{t}\Psi_{R})]=[H,\rho_c]_{+}
					\label{eq:anticom2}
\end{eqnarray}
As far as I am aware this equation has not appeared in this form in the literature before it was introduced by Brown and Hiley \cite{mbbh00}, but was hinted at in Carruthers and Zachariasen \cite{cz83}.  

\noindent Using $H=p^{2}/2m+V$ for the Hamiltonian, it is straight forward to show equation \eqref{eq:conprob2} becomes
\begin{eqnarray*}
\partial_{t}P+\nabla(P\nabla S/m)=0
\end{eqnarray*}
which is clearly the usual equation for the conservation of probability.

Equation \eqref{eq:anticom2} can be considerably simplified if we write $\Psi_{L}=R\exp[eS]$, where $e$ is the generator of the Clifford algebra $C_{(0,1)}$ \cite{bhbc10}.    After some work we find the equation reduces to the quantum Hamilton-Jacobi equation
\begin{eqnarray}
\partial_{t}S+(\nabla S)^{2}/2m +Q + V =0		\label{eq:QHJ}
\end{eqnarray}
where $Q=-\nabla^{2}R/2mR$ is the usual expression for the quantum potential \cite{bh93}.  In this way we have derived the quantum Hamilton-Jacobi equation for a Schr\"{o}dinger particle from the quantum algebra itself.  Notice there is no appeal to classical mechanics to arrive at this equation.  In fact 
the  two dynamical equations (\ref{eq:conprob2}) and (\ref{eq:anticom2}) form the basis of the generalised Bohm approach to quantum mechanics.  By applying these equations to the Pauli and Dirac Clifford algebras, one can obtain a complete description of the quantum dynamical equations for the Pauli and Dirac particles. 

Before we can do this we have to know how the Bohm momentum and the Bohm energy can be expressed in their algebraic form.  A detailed discussion of these generalisations using the energy-momentum tensor can be found in Hiley and Callaghan \cite{bhbc10} \cite{bhbc10a}.  Here we will be content to simply write down the expressions.  For the Pauli and Dirac particles we have
\begin{eqnarray}
\rho P^{j}(t)=-i\Psi_{L}\overleftrightarrow\Delta^{j}\widehat\Psi_{L}=-i\left [(\overrightarrow\Delta^{j}\Psi_{L}){\widehat\Psi_{L}}-\Phi_{L}({\widehat\Psi_{L}}\overleftarrow\Delta^{j})\right]								
			\label{eq:P}					
\end{eqnarray}
and an energy by
\begin{eqnarray}
\rho E(t)=i\Psi_{L}\overleftrightarrow\Delta^{0}\widehat\Psi_{L}=i[(\overrightarrow\Delta^{0}\Psi_{L}){\widehat\Psi_{L}}-\Psi_{L}({\widehat\Psi_{L}}\overleftarrow\Delta^{0})]
			\label{eq:E}					
\end{eqnarray}

By using the derivatives listed above we can obtain expressions for the respective $P_B$ and $E_B$.  These can then be used in equation (\ref{eq:anticom2}) to obtain the respective quantum Hamilton-Jacobi equations.  The algebraic details to arrive at the equations can be rather tedious so will not be discussed here.  The  full treatment of the Pauli and Dirac particles will be found in Hiley and Callaghan  \cite{bhbc12} \cite{bhbc10} and \cite{bhbc10a}.

The method we have outlined in this paper improves considerably on the previous attempts to extend the Bohm approach to spin and to the relativistic domain.  For example, as we have already remarked, 
the Pauli equation has already been treated using Euler angles by Bohm, Schiller, and Tiomno, \cite{dbst55} (See also Dewdney {\em et al} \cite{cdjv88}.)   What they did was to express the spinor in terms of Euler angles so that 
\begin{eqnarray*}\Psi=\begin{pmatrix}
      \psi_{1}    \\
      \psi_{2}  
\end{pmatrix}=
\begin{pmatrix}
     \cos(\theta/2) \exp(i\phi/2)    \\
      i\sin(\theta/2) \exp(-i\phi/2)
\end{pmatrix}\exp(i\psi/2)
\end{eqnarray*}
Here we see the azimuthal angle, $\psi$, appearing in a position that enabled it to be identified as a `common phase', although no justification was given.
Substituting this into equation (\ref{eq:P}) above, we find the Bohm momentum for the Pauli particle is
\begin{eqnarray}
 P_{B}=(\nabla S+\cos\theta\nabla\phi)/2		\label{eq:paulipEul}
\end{eqnarray}
This expression was first obtained in Bohm, Schiller and Tiomno [BST] \cite{dbst55}.  On the other hand using the expression of the element of the minimal left ideal gives, when converted into wave functions,
\begin{eqnarray}
2\rho  P_{B}(t)=i[(\nabla\psi_{1})\psi^{*}_{1}-(\nabla\psi^{*}_{1})\psi_{1}+(\nabla\psi_{2})\psi^{*}_{2}-(\nabla\psi^{*}_{2})\psi_{2}]		\label{eq:PB}	\end{eqnarray}
which when converted into Euler angles produces exactly equation (\ref{eq:paulipEul}).
Writing $\psi_{1}=R_{1}e^{iS_{1}}$ and $\psi_{2}=R_{2}e^{iS_{2}}$, equation \eqref{eq:P}  becomes
\begin{eqnarray}
\rho  P_{B}(t)=(\nabla S_{1})\rho_{1}+(\nabla S_{2})\rho_{2}	\label{eq:PB2}
\end{eqnarray}							
where $\rho_{i}=R_{i}^{2}$. The meaning becomes more transparent if we write $ P_{i}=\nabla S_{i}$ when the expression for the momentum becomes
\begin{eqnarray}
\rho  P_{B}(t)= P_{1}\rho_{1} +  P_{2}\rho_{2}		
				\label{eq:MP}			
\end{eqnarray}
Thus we see that in terms of the usual approach $P_{B}(t)$ is the weighted mean of the momentum that can be attributed to each component of the spinor acting by itself.  This result was already noted in Bohm and Hiley \cite{bh93}.

Similarly the energy expressed in equation (\ref{eq:E}) becomes
\begin{eqnarray}
\rho E_{B}(t)=-[(\partial_{t}S_{1})\rho_{1}+(\partial_{t}S_{2})\rho_{2}].	\label{eq:E2}
\end{eqnarray}	
When this equation is converted into Euler angles it becomes
\begin{eqnarray}
E_{B}(t)=	-(\partial_{t} S +\cos \theta\partial_{t}\phi)/2
\end{eqnarray}
which agrees with the expression found in BST \cite{dbst55}.

Again using the polar form for each spinor component $\psi_{i}=R_{i}e^{iS_{i}}$, the energy can be written in the form
\begin{eqnarray}
\rho E_{B}=E_{1}\rho_{1}+E_{2}\rho_{2}	
				\label{eq:ME}		
\end{eqnarray}
which is clearly seen as the weighted mean of the energy associated with each component of the spinor.

It should  be noted that Hestenes and Gurtler, \cite{dhrg71} also have a Clifford algebra treatment of spin, but they do not discuss its relevance to the Bohm approach.  Doran and Lasenby  also obtain an expression for the `guidance condition' \cite{cdal93} but do not discuss the quantum Hamilton-Jacobi equation.  However
 the Dirac theory has only been partially treated previously in terms of the Bohm approach by Bohm and Hiley \cite{bh93} and by Hestenes \cite{dh03} where he mentions a possible connection with Bohm's quantum potential.

 \section{Generalised Phase Space.}
 
 In the last section we report the work done extending and clarifying the role of orthogonal Clifford algebras in the Bohm approach.  This leaves the question of how the symplectic structure enters the discussion.  Bohm and I began exploring the role of the symplectic structure underlying both classical and quantum systems  in our paper ``On a Quantum Algebraic Approach to a Generalized Phase Space'' \cite{dbbh82a}.  Our work was not based on an entirely new idea since some features of our approach were already contained in the Moyal algebra \cite{jm49} which in turn accommodates the Wigner distribution used in what is termed, wrongly in my view, the `semi-classical' approach to many-body  problems \cite{cz83}.
   
What we showed in our paper was that  by considering two points in configuration space, we could construct a phase space description of quantum phenomena, a feature that many physicists think is impossible, or at least limited in some way, because of the uncertainty principle.  However a careful study of the formalism shows that the $x$ and $p$  used in this approach are not the $(x, p)$ coordinates of the particle, but rather the mean coordinates of an extended structure or `blob' \cite{mg10} in a non-commutative phase space.  This gives mathematical form to Weyl's notion of a particle  
\begin{quote}
Hence a particle itself is not even a point in field space, it is nothing spatial (extended) at all. However, it is confined to a spatial neighbourhood, from which its field effects originate \cite{hw24}.
\end{quote}
Thus our idea that the particle is a quasi-local invariant feature of the total process receives support from Weyl, whose algebraic approach has strongly influenced my own thinking.

An important feature of this approach is that
in the limit of order $\hbar$, the algebraic structure reduces to the Poisson algebra of classical phase space and the `blob' becomes a point particle \cite{bh13}.

 To return to our original paper \cite{dbbh82a}, it should be noted that we did not pay sufficient attention to the non-commutative nature of this phase space.  Rather we proposed a way of defining a modified Wigner distribution that would always provide a positive probability distribution.  This approach would predict differences from standard quantum mechanics, but I now believe that our attempt to construct a positive-definite  distribution was somewhat beside the point. The reason being that the Wigner distribution is actually the density matrix in disguise, being expressed in terms of the sum-and-difference of coordinates and there is no reason why such a density matrix should always be positive. 

 There we also attempted to extend our ideas to the Dirac theory \cite{dbbh83}, but again the attempt was rather premature for two reasons: (1) we did not yet have a sufficient understanding  of how to apply the orthogonal Clifford algebra to the Dirac equation itself in the context of the Bohm approach and (2) we did not have a sufficient understanding of the symplectic Clifford algebra to complete the discussion.  As we have already seen in the previous section, the first difficulty has now been removed by Hiley and Callaghan \cite{bhbc10a}.  We are therefore now in a position to go on to discuss how the symplectic Clifford algebra enters the discussion.

\subsection{Moyal and von Neumann Algebras.}

To explain the role of the symplectic Clifford algebra, we must return to consider two very early papers.  The first is a long forgotten paper by von Neumann \cite{vn31}, where he proposes a purely algebraic approach to the quantum formalism.  Here he emphasised the structure of the operator algebra rather than its representation operating on vectors in Hilbert space \cite{bhbc12}.  In other words, the wave function plays a minor role, as all the necessary information needed to describe the quantum state is already contained in the operator algebra itself as we have  already pointed out \cite{ffbh80a}.
 The second appeared two decades later.  It is the classic paper by Moyal \cite{jm49}.  In this paper he shows how the operator formalism can be regarded as providing a generalisation of ordinary statistical methods applied to a non-commutative phase space.
  
  It turns out that the mathematical structure proposed by Moyal is isomorphic to that described by von Neumann.  The von Neumann structure depends on two parameters $\alpha$ and $\beta$, but no physical meaning is attributed to them.  On the other hand  Moyal calls them $p$ and $x$ giving them the meaning of the momentum and position of a particle.  In this way he draws attention to the possibility of describing quantum phenomena in a phase space.  However we now know this is no classical phase space, it is a non-commutative phase space, where the symbols $p$ and $x$ are subject to star-multiplication $f(p,x)\star g(p,x)$, which is non-commutative \cite{zfc06} \cite{bh13}.  This product is called the Moyal product even though it was actually defined in the original von Neumann paper. Furthermore as we have noted above, the coordinates $x$ and $p$ are not the coordinates of a point particle, but the mean values of a quantum `blob' in a symplectic space which becomes the classical phase space in the classical limit.

What this result means is that the Moyal algebra should not be regarded as `semi-classical' as it is at the heart of the standard quantum formalism, which is, of course,  why the approach reproduces exactly all the quantum expectation values.  It is a way of treating quantum phenomena in a non-commutative space, which for obvious reasons, we have called  a `non-commutative phase space'.  We will emphasise again, the great advantage of this approach is that to $O(\hbar)$ the `blob' becomes a point and the $p$ and $x$ now refer to the momentum and position of a classical point particle.  Furthermore the Moyal algebra becomes the Poisson algebra in this limit so that the resulting phase space is commutative, being the phase space of classical mechanics.

Once again we must ask, ``What has this all to do with Bohm's model?" In actual fact I initially put the question to myself in another way, ``If Moyal's work produces the correct quantum expectation values based on a phase space, and the Bohm model, based on what appears to be a different phase space, produces the correct quantum expectation values, then surely there must be some relation between the two sets of formalism.  What is this relation?" 

 A close examination of the Appendix to the Moyal paper, revealed a derivation of  the so-called `guidance' condition $p=\nabla S$ directly from the Moyal algebra itself with no appeal to any classical limit.  Moyal then developed the transport equation for this momentum and found it was identical to the Bohm quantum Hamilton-Jacobi equation with the quantum potential clearly visible \cite{bh10b}.  Moyal did not draw attention to this term since his work pre-dates Bohm's, rather he simply showed how the Schr\"{o}dinger equation could be obtained from the equation.  Thus  two very different starting points, Moyal's and Bohm's, led to identical equations \cite{bh10b}.  This fact alone again points to a deep connection but we need to find exactly what this connection is.

\subsection{Relation between Bohm and the Moyal Algebra.}

What is emerging here is that if we start, as von Neumann does,  by examining the Heisenberg group algebra, we are led to a non-commutative symplectic geometric algebra, an analogue to the orthogonal Clifford geometric algebra.  Now the symplectic geometry is a common symmetry in both the quantum and classical domains, the geometry being non-commutative in the quantum domain but becoming  the Poisson algebra with commutative products in the classical domain.  This is a particular example of a deformation algebra, a simple account with further references will be found in Hirshfeld and Henselder \cite{ahph01}.

As we have already seen,  the {\em algebraic density element}  $\rho_\psi$ is key to the Clifford algebraic approach.  It may seem odd to introduce a distribution function to describe a single `particle', but it should be remembered that in a process based approach, the notion of a `particle' is not a sharply defined point object but a quasi-local invariant feature of the total process.  If we try to replace this quasi-local by a point particle, the position of the point will be ambiguous.  Nevertheless we can characterise each individual process by two parameters which can be regarded as the mean position and the mean momentum.  It is this feature that allows us to track an individual process in space-time.  In the classical limit these variables can be associated with a localised energy, the point particle.

To find expectation values of quantum operators, now treated as elements of the algebra, we must first find the corresponding Weyl symbol,
\begin{eqnarray*}
\hat A \leftrightarrow A(p,x)
\end{eqnarray*}
 For more details of how this is actually accomplished,  see  de Gosson  \cite{mdg10}.  The well known ambiguity in this symbol is removed in our approach by insisting on symplectic covariance \cite{mgbh11}.  Once we have the expression of the Weyl symbol, the expectation value of the operator $\hat A$ is calculated  from
 \begin{eqnarray*}
 \langle \hat A\rangle=\int\int A(p,x)F_\psi(p,x)dpdx
 \end{eqnarray*}
 This procedure always produces exactly the same expectation value as  calculated from the standard formalism.
 
 Given a distribution which depends on two sets of random variables, we also have the possibility of determining conditional expectational values [CEV].  For example, for the single particle, we can ask for the CEV of the momentum by treating  $F_\psi(p,x)$ as a probability distribution\footnote{Note $F_\psi(p,x)$ is not strictly a probability distribution as it can become negative. Both Bartlett \cite{mb44} and Feynman \cite{rf87} have shown how these probabilities can be justified. However the probability distribution is just the density matrix expressed in terms of a mean position and a mean momentum of an extended process.  We then find, for example, some conditional mean-square deviations take on negative values \cite{w05}.  This is why Moyal \cite{jm49} regarded his theory as a generalised statistical theory.}.  This CEV is given by
 \begin{eqnarray}
 {\bar p}_\psi=\int p F_\psi(p,x)dp	\label{eq:cevp}
 \end{eqnarray}
 If we apply this to a non-relativistic spin-less particle, usually described by a wave function, which we write as $\psi=Re^{iS/\hbar}$, we find that
 \begin{eqnarray*}
 {\bar p}=\nabla S=P_B
  \end{eqnarray*}
which is just the Bohm momentum $P_B$.  Just for the record, it should be noted that this method of deriving $P_B$ first appeared in the appendix of Moyal's paper referred to above \cite{jm49}.

To find the Bohm energy, we simply extend the phase space to include the energy and time so that the distribution function becomes $F_\psi(p, x, E, t)$.  Using this distribution, the conditional expectation value of the energy is given by
\begin{eqnarray*}
\bar E (x,t)=\int E F_\psi(p, x, E, t)dp.
\end{eqnarray*}
We find
 \begin{eqnarray*}
\bar E (x,t) =-\frac{\partial S}{\partial t} = E_B.
\end{eqnarray*}

As is well known we can find the expectation value of any element $\hat A$ from conditional values via
\begin{eqnarray*}
\langle \hat A\rangle=\int P(x,t)P(A|x)dx
\end{eqnarray*}
so that, for the momentum
\begin{eqnarray*}
\langle \hat P\rangle=\int P(x,t) P(p|x)dx.
\end{eqnarray*}

Above we have shown one way of finding the CEV, but there is another somewhat surprising way that has recently emerged.  

The CEV of $\hat P$ in the quantum formalism can be written in the form
\begin{eqnarray*}
P(p|x)=\frac{\langle x|\hat P|\psi(x)\rangle}{\psi(x)}
\end{eqnarray*}
In the quantum literature this is called a weak value of the momentum \cite{rl05}.  It is easy to show that by using $\psi=Re^{iS/\hbar}$, the weak value for the momentum is 
\begin{eqnarray*}
\frac{\langle x|\hat P|\psi(x,t)\rangle}{\psi(x,t)}=\nabla S,
\end{eqnarray*} 
which is, of course, just the Bohm momentum.  Now it is well known that a CEV can be measured.  Thus the Bohm momentum, being a CEV of the momentum, must be measurable.  Indeed it has actually been measured in an interference experiment by Kocis {\em et al.} \cite{skms11}.  Flack {\em et al} \cite{rfbh13} are in the process of measuring the Bohm momentum of aluminium atoms.

By now it should be clear that the Bohm model deals with conditional probabilities within the standard quantum formalism.  If we merely limit ourselves to the standard Hilbert space formalism, it is easy to miss this deeper structure.  This position becomes even more compelling if we extend this formalism to the case of spin and to relativistic particles like the Dirac electron.  In the next section we sketch how the formalism can be extended into these domains.

Before moving to consider this generalisation, we want to emphasise again that within our approach the `particle' is not a `rock-like' point particle.  It has an extended structure which we have called the `quantum blob'.  However this name, although suggestive, should be replaced by a more formal concept. Maurice de Gosson \cite{mg01} has suggested that it should be called a `metatron'.  This suggestion stems from the mathematical structure in which it appears.  

The symplectic group, like the orthogonal group, has a double cover, the metaplectic group.  By lifting the classical motion from the symplectic manifold to the covering metaplectic manifold, we find the appearance of a wave function and the Schr\"{o}dinger equation \cite{vgss}.  Thus any quasi-local semi-autonomous structure will unfold under the metaplectic group and for this reason we call this `object' a `metatron'.  What we find is that this metatron is an individual entity described by its mean momentum and mean position. 

As we have seen the Bohm model suggests that these parameters identify a particular process within an ensemble of possible individual processes, so we propose to regard each individual process as the evolution of a metatron. Thus we can identify each individual process by a pair of parameters enabling us to distinguish each individual process in the ensemble.  Bohm noticed a similar parameterisation of each process but, for simplicity, called the individual a `particle'.  The term metatron is a better name the individual process unfolds via the metaplectic group.   In the classical limit when terms involving  $\hbar^2$ and above can be neglected, the metatron become the classical point particle subjected to Hamiltonian dynamics.

\subsection{The Moyal Algebra with Spin}

In the previous subsection we showed how the Bohm model emerged from the von Neumann algebra when expressed in the language of the Moyal algebra. However the Moyal algebra does not include a description of spin, nor is it relativistic.  In this section we will show how to combine the Moyal algebra with the orthogonal Clifford algebra so as to include spin and even relativity.  

As explained above,  we introduced a Clifford density element $\rho_\psi$ which we wrote as
\begin{eqnarray*}
\rho_\psi(x,t)=\Psi_L(x,t)\widehat\Psi_L(x,t)
\end{eqnarray*}
 where $\Psi_L(x,t)$ is an element of a suitable chosen minimal left ideal.

 In order to construct a phase space, we follow the procedure developed in Bohm and Hiley \cite{dbbh82a}  where we formed a two-point density function
 \begin{eqnarray*}
\rho_{\Phi}(x_1,x_2,t)=\Phi_L(x_1,t)\widehat\Phi_L(x_2,t)
\end{eqnarray*}
This means that we can write a generalised Wigner-Moyal distribution function as
\begin{eqnarray*}
F_\Phi(P,X,t)=\int Tr[\Phi_L(x_1,t)\widehat \Phi_L(x_2,t)e^{ip.y}]d^3y
\end{eqnarray*}
where $X=(x_2+x_1)/2$ and $y=x_2-x_1$.  
This distribution was used in Hiley \cite{bh13} to show that, after some work, the conditional expectation value of the momentum $\widehat P$ is 
\begin{eqnarray}
\rho(x,t)P_B(x,t)=\int p F_\psi(p,x,t)=\rho_1(x,t)\partial_xS_1(x,t) +\rho_2(x,t)\partial_xS_2(x,t)		\label{eq:paulip}
\end{eqnarray}
which is exactly the equation (\ref{eq:PB2}).  By generalising the Wigner distribution to include energy, we find the conditional expectation value for the energy is exactly that obtained in equation (\ref{eq:E2}).  Once again we see the intimate relation between the the Moyal algebra and the generalised Bohm approach.

 
 \subsection{Non-commutative Phase Spaces}
 
 In the previous subsection, we introduced a pair of time development equations for the Clifford density element $\rho_\psi$.  In Hiley \cite{bh13} it was shown how the Clifford algebra approach could be generalised to apply to the non-commutative phase space introduced by Moyal.  When this is done one again finds two time development equations.  The analogue of equation (\ref{eq:conprob2}) takes the form
\begin{eqnarray}
\frac{\partial F_\psi(p,x,t)}{\partial t}+\{F_\psi(p,x,t),H(p,x)\}_{MB}=0	\label{eq:LE}		
\end{eqnarray}
where $\{F_\psi(p,x,t),H(p,x)\}_{MB}$ is the Moyal bracket defined by
 \begin{eqnarray*}
\{a,b\}_{MB}=\frac{a\star b-b\star a}{i\hbar}.
\end{eqnarray*}
Thus we see that the commutator bracket in equation (\ref{eq:conprob2}) has been replaced by the Moyal bracket.

The analogue of the second equation (\ref{eq:anticom2})  is 
\begin{eqnarray}
{\cal E}(p,x,t)+\{H,F\}_{BB}=0 	\label{eq:MQHJ}	
\end{eqnarray}
where the anticommutator has been replaced by $\{H,F\}_{BB}$, the Baker bracket \cite{bak}  (or Jordan product) defined by
\begin{eqnarray*}
\{a,b\}_{BB}=\frac{a\star b+b\star a}{2}.
\end{eqnarray*}
The left hand side of equation (\ref{eq:anticom2}) has been replaced by ${\cal E}(p,x,t)$, which  is defined by
\begin{eqnarray*}
{\cal E}(p,x,t)=-i(2\pi)^{-1}\int e^{-i\tau p}\left[\psi^*(x-\tau/2)\overleftrightarrow\partial_t\psi(x+\tau/2)\right]d\tau,
\end{eqnarray*}

The $\star$-product can, for our purposes, be defined as
\begin{eqnarray*}
a(x,p)\star b(x,p)=a(x,p)\exp[i\hbar(\overleftarrow\partial_x\overrightarrow\partial_p-\overrightarrow\partial_x\overleftarrow\partial_p)/2]b(x,p)
\end{eqnarray*}
which shows that the Moyal bracket will be a power series in $\hbar$.  If we  retain only the terms to $O(\hbar)$, we find 
\begin{eqnarray*}
\mbox{Moyal bracket}\rightarrow \mbox{Possion bracket}.
\end{eqnarray*}
 While in the case of the Baker bracket we find
\begin{eqnarray*}
\mbox{Baker bracket}\rightarrow \mbox{simple commutative product}.
\end{eqnarray*}
I want to emphasise here that it is the classical limit of the Baker bracket that reduces the Moyal algebra to the commutative algebra.  The Moyal bracket, although playing the role of the commutator in equation (\ref{eq:LE}), does not vanish in the classical limit.  Rather it becomes the Poisson bracket, forming the basis of classical mechanics.  More details of the Moyal algebra can be found in Zachos {et al.} \cite{zfc06}

Indeed in this limit we find that equation (\ref{eq:conprob2}) becomes the classical Liouville equation, while equation (\ref{eq:anticom2}) becomes the classical Hamilton-Jacobi equation.  This last result justifies  the name `quantum Hamilton-Jacobi equation' for equation (\ref{eq:QHJ}).  Notice that the quantum energy plays a significant role in this first order differential equation and vanishes in the classical limit. 

The most important lesson we learn from the Moyal approach is that the  quantum algebra contains classical mechanics as a limiting case.  There is no need to look for a correspondence between  commutator brackets and Poisson brackets, a process which fails, as was demonstrated in the well known Groenwald-van Hove  `no-go' theorem \cite{vgss}.  Nor is there any need to introduce decoherence to reach the classical level.

 \subsection{Shadow Phase Spaces and the Explicate Order.}
 
 Finally I would like to link the progress that has been made with the formalism discussed above and the more philosophical ideas that David Bohm felt would provide the background to give a more intuitive understanding of the quantum formalism \cite{db80}. 
 
 To do this let us first return to a technical point.   We introduced in equation (\ref{eq:cevp}) the CEV of the momentum.  Notice, however we could have alternatively introduced the CEV of a position variable $x$.  We could then define 
 \begin{eqnarray}
 \bar X=\int x F_\psi(p,x)dx	\label{eq:cevx}
 \end{eqnarray}
 It can be shown that in this case  
 \begin{eqnarray*}
  \bar X(p)=-\frac{\partial S_p}{\partial p}
 \end{eqnarray*}
 This  move is equivalent of going to the $p$-representation.  Recall that in this representation, the operator $\hat x$ is given by $\hat x =i\frac{\partial}{\partial p}$. Thus the Moyal phase space contains both of these representations, retaining the symmetry found in the standard formalism.
 
 The use of conditional expectation values produces a new situation.  Just  as we treated the pair  $(\bar p,x)$ as defining what I will call a Bohmian phase space, we can use the pair  $(p,\bar x)$ as defining another Bohmian phase space.  In Brown and Hiley \cite{mbbh00} we explored the structure of these two phase spaces and found that the same quantum system produced {\em two different} sets of trajectories.  Further investigations by Brown \cite{mb04}, using fractional Fourier transformations, showed the existence of other possibilities, so the conclusion was that the Bohm approach has the possibility of at least two phase spaces, a $(\bar p,x)$ phase space or a $(p,\bar x)$ phase space. On the positive side this restores the mathematical symmetry of the $x$-representation and the $p$-representation that Heisenberg \cite{wh58} complained was missing in the original Bohm proposal \cite{bo52}.  However by doing this we have returned to the situation that Pauli \cite{wp79} complained about, namely,
 \begin{quote}
One can look at the world with the $p$-eye and with the $x$-eye, but if one wishes to open both eyes at the same time, one goes wrong.
 \end{quote}
 Thus the attempt to produce one unambiguous representation of quantum phenomena {\em seems} to have been thwarted once again.  
 
 However the appearance of dual images are not totally unknown.  Indeed ambiguous images such as the pelican-antelope (see Figure 1) or the old lady-young lady are by now well known.  
 
 \begin{figure}[h] 
    \centering
    \includegraphics[width=1in]{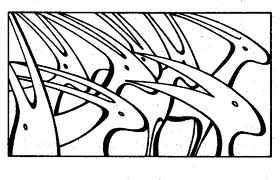} 
    \caption{Pelican-Antelope Illusion.}
    \label{fig:example}
 \end{figure}
 \noindent Here the ambiguous structure of lines can be given two meanings which appear to be contradictory, the contradiction arises because we try to fit the structure to an image of  some  previously experienced image, not of what is actually there.  
 
 In quantum domain we are trying to fit classical images to the phenomena and failing.  This ambiguity  arises, not because there is no underlying process, but because the process is different from what we expected.  What underlying structure could possibly lead to this ambiguous situation? 
 
  For us, and also for Bohr, the most important new feature of quantum phenomena is the notion of {\em unbroken wholeness}.  We, together with our instruments, are inside the world looking out.  We are not gods outside looking in.  Being `inside' means that our attempts to construct a `third-person  view, an explicate order, must necessarily produce a partial or, if you like, ambiguous account.   But this does not mean there is no underlying structure which can account for the partial images.  What we should be doing is to try to understand this underlying order.
  
In order to achieve this understanding we have to develop a radically new way of thinking that makes it clear what limitations to which we are subjected, not in an absolute sense, but in a sense that we can only have a set of partial views--  a set of explicate orders.  These are not any old orders, since they must cohere within a deeper order.  This deeper order is the implicate order.   This is the essence of Bohm's notion of the implicate order \cite{db80}.
 
 A sketch of the informal ideas with which to understand  quantum phenomena is already contained in Bohm's discussion of the {\em process of becoming} \cite{bo57d}.  I will not try to repeat Bohm's arguments here as they are beautifully presented in his extensive writings.  What I want to draw attention to here is the fact  that already in the earlier writings of Grassmann \cite{hg94}, Hamiltion \cite{wh67}, and Clifford \cite{wc82}, attention had been drawn to the possibility of an underlying basic process  which  could be adequately discussed in terms of an algebra.  Not an algebra in space-time, which is what I have been exploiting above, but algebras from which space and time is abstracted, which is what still has to be done. We can begin to get an idea of where to start by following Liebnitz 
 \begin{itemize}
\item
 Time is the order of succession $\Rightarrow$ multiplication. 
 \item
 Space is the order of coexistence $\Rightarrow$ addition.
 \end{itemize}
 which implies an algebra since two binary relations define an algebra  
 
Since the process of becoming is basic, our algebra is basic.  Now we have to abstract space and time  from this algebra.  How do we do this?  The answer lies in the ideas developed by Gel`fand \cite{jd97}.  There are essentially two approaches to a dynamical system.  The traditional way, which assumes an {\em a priori } given manifold with topology and metric on which one can discuss the time development of a system placed in that space in terms of some algebraic structure.  

Alternatively, we can start with the dynamical algebra and, provided the algebra is commutative, one can deduce all the properties of the underlying manifold from the algebra itself.  In other words the nature of the space-time properties are encoded in the algebra.  This only works for commutative structures.  If the algebra is non-commutative, as it is for quantum phenomena, then there is no unique underlying manifold.  All you can find are sets of `shadow manifolds' or, if you like, explicate orders.  This is exactly what we find here.  For a more detailed discussion see Hiley \cite{bh11}.
 
 Note, however these are not sets of arbitrary explicate orders.  Each one takes it form from the same algebra, the same implicate order.  In quantum mechanics we know what defines each explicate order--the experimental conditions.  The experimental conditions impose constraints on the overall order so that we now open up the way to provide a coherent way of understanding quantum phenomena in terms of a real underlying process, a process that cannot be uniquely described in a single space-time order.
 
 \section{Conclusion.}
 
 In this paper we have reviewed what I will call the Bohm programme aimed at providing a model of an independent reality that lies behind the quantum formalism.  We have taken his physical  and philosophical insights and developed the mathematics in which these ideas can be given a precise form.  In this paper we have concentrated on low energies (below pair production energies) where a single particle can be discussed both at the non-relativistic and relativistic levels.
 
 The mathematics is not new, being based on an  algebra first introduced by von Neumann \cite{vn31} and later developed by Moyal \cite{jm49}.  This algebra, when supplemented by the orthogonal Clifford algebra \cite{bhbc12}, gives a complete account of single particle quantum phenomena.  The extension of these techniques many-body systems \cite{bhnonl} and to quantum field theory \cite{dbbhpk} \cite{bh93} will not be discussed here, but will be presented elsewhere.
 
 Notice here we are essentially exploiting the structure of a von Neumann algebra of type I.  Type II and III von Neumann algebras have a much richer structure that enables us to link quantum theory with thermodynamics \cite{ge72} \cite{ac90}.  This opens up a very rich area for exploration which, again, we leave for another paper.
 
 Although we have shown that the algebraic formalism contains both quantum and classical dynamics in one mathematical structure,  big questions remain:  How is it that, in our macroscopic world, there is a unique space-time order?  What are the principles that lead from this process description to the appearance of the quasi-local semi-autonomous particle structures we find around us?  How does the space-time structure emerge from this underlying process?
 
 What is still missing is a general principle which ensures that stable systems can emerge.  At present all we have some rather vague ideas as to how this can happen, but as yet it is not clear enough to be discussed here. This is a question for the future. 
 
\section{Acknowlegements.}

This work emerges from many valuable discussions I have had with many people that I cannot hope to thank everybody who has helped.  Nevertheless I would like to thank Maurice de Gosson and Ernst Binz for  their mathematical input, without which none of this would have been possible.  I should also like to thank Ray Brummelhuis and members of the TPRU for their patience in our many discussions.


\begin{thebibliography}{99}

\bibitem{aa} Aspect A.,  Dalibard J. and  Roger G.,  Experimental Realization of Einstein-Podolsky-Rosen-Bohm Gedankenexperiment: A New Violation of Bell's Inequalities, {\em Phys. Rev. Letts.}, {\bf 49},  (1982), 91-94.

\bibitem{bak}   Baker, G. A., Jr., Formulation of Quantum Mechanics Based on the Quasi-Probability Distribution Induced on Phase Space, {\em Phys. Rev.}, {\bf 109}, (1958), 2198-2206.

\bibitem{mb44} Bartlett, M. S., Negative Probability, {\em Math. Proc. Cam. Phil. Soc}. {\bf 41}, (1945) 71-3.

\bibitem{jb87a}   Bell, J. S., {\em Speakable and Unspeakable in Quantum Mechanics}, p. 163,  Cambridge University Press, Cambridge, 1987.

\bibitem{bo51}Bohm, D., {\em Quantum Theory},  Prentice-Hall, Englewood Cliffs, N.J.  (1951)

\bibitem{bo51a} Bohm, D., {\em Quantum Theory}, p. 167, Prentice-Hall, Englewood Cliffs, N.J.  (1951).

\bibitem{bo52} Bohm, D.,  A Suggested Interpretation of the Quantum Theory in Terms of Hidden Variables, I, {\em Phys. Rev.}, {\bf 85}, (1952), 166-179, and II, {\em Phys. Rev.}, {\bf 85}, (1952), 180-193.

\bibitem{db53} Bohm, D., Comments on a Letter Concerning the Causal Interpretation of Quantum Theory, {\em Phys. Rev}. {\bf 89}, (1953) 319-20. 

\bibitem{dbst55} Bohm, D., Schiller, R. and Tiomno, J., A Causal Interpretation of the Pauli Equation (A) and (B), {\em Nuovo Cim}. Supp. {\bf 1}, (1955) 48-66 and 67-91.

\bibitem{bo57} Bohm, D., {\em Causality and Chance in Modern Physics}, Routledge \& Kegan Paul, London, 1957.

\bibitem{bo57b} ibid, p. 143.
\bibitem{bo57c} ibid, p. 145.
\bibitem{bo57d} ibid, p. 147.

\bibitem{bo62} Bohm, D. in {\em Radiation and High Energy Physics}, ed.  Bates, D. R., p. 345, Academic Press, 1962.

\bibitem{dbsb} Bohm, D.,  Problems in the Basic Concepts of Physics,  In {\em Satyendranath Bose 70th Birthday Commemoration Volume},  Part II, 279-318. Calcutta,  1965.

\bibitem{dbkyo} Bohm, D., Space, Time and the Quantum Theory Understood in Terms of Discrete Process, {\em Proc. of the Int. Conf. on Elementary Particles}, Kyoto, Japan, 252-286, (1965).


\bibitem{bh70} Bohm, D., Hiley, B. J. and Stuart, A. E. G.,  On a New Mode of Description in Physics, {\em Int. J. Theor. Phys.}, {\bf 3}, (1970), 171-183.

\bibitem{db71} Bohm, D., Quantum Theory as an Indication of a New Order in Physics Part A: The Development of New Orders as shown Through the History of Physics, {\em Found. Phys}. {\bf 1}, (1971), 359-81.

\bibitem{db73} Bohm, D., Quantum Theory as an Indication of a New Order in Physics Part B: Implicate and explicate order in physical law,  {\em Found. Phys.}, {\bf 3}, (1973), 139-68. 

\bibitem{bhnonl} Bohm, D., and Hiley, B. J.,  On the Intuitive Understanding of Nonlocality as Implied by Quantum Theory,   {\em Found. Phys.},  {\bf 5}, (1975), 93-109.


\bibitem{jlmp} Bohm, D., and Hiley, B. J., On the Intuitive Understanding of Nonlocality as Implied by Quantum Theory, in {\em Quantum Mechanics, a Half Century Later}, ed. Lopes, J. L. and Paty, M., pp, 207-25, Reidel, Dordrecht 1975.


\bibitem{db80}  Bohm, D., {\em Wholeness and the Implicate Order}, Routledge, London, 1980.


\bibitem{dbbh84}  Bohm, D. and Hiley, B. J., Generalization of the Twistor to Clifford Algebras as a Basis for Geometry,   {\em Revista Brasilera de Fisica, Vol. Especial Os 70 anos de Mario Sch\"{o}nberg}, 1-26, (1984).

\bibitem{db86} Bohm, D., Time, the Implicate Order and Pre-Space, in {\em Physics and the Ultimate Significance of Time}, ed. D.R. Griffin, 172-6 \& 177-208, SUNY Press, N.Y., 1986.

\bibitem{dbdp87} Bohm, D., and Peat, F. D. {\em Science, Order and Creativity}, Routledge, London, 1987.

\bibitem{db88}   Bohm, D., A Realist View of Quantum theory in {\em Microphysical Reality and Quantum Formalism}, ed. van der Merwe, A. et al., p. 3, Kluwer, 1988.

\bibitem{dbbh82a}  Bohm, D. and Hiley, B.J.  On a Quantum Algebraic Approach to Generalised Phase Space,   Found. Phys. 11, (1981) 179-203. 

\bibitem{dbbh82} Bohm, D., and Hiley, B. J.,  The de Broglie Pilot Wave Theory and the Further Development of New Insights Arising out of It, {\em Found. of Phys}., {\bf 12}, (1982) 1001-16. 

\bibitem{dbbh83}   Bohm,  D. and Hiley, B. J., Relativistic Phase Space Arising out of the Dirac Algebra,  {\em  Old and New Questions in Physics, Cosmology, Philosophy and Theoretical Biology}, ed. A. van der Merwe, 67-76, Plenum Press, New York, 1983.

\bibitem{dbcdbh} Bohm D., Hiley, B. J., and Dewdney, C.,   A Quantum Potential Approach to the Wheeler Delayed-Choice Experiment,    {\em Nature}, {\bf 315}, (1985), 294-297.

\bibitem{bh87} Bohm, D., and Hiley, B. J.,  An Ontological Basis for Quantum Theory: I - Non-relativistic Particle Systems,  {\em Phys. Reports} {\bf 144}, (1987), 323-348.

\bibitem{dbbhpk}   Bohm, D.,  Hiley, B. J. and  Kaloyerou, P. N., An Ontological Basis for the Quantum Theory: II -A Causal Interpretation of Quantum Fields, {\em Phys. Reports},  {\bf 144}, (1987) 349-375.

\bibitem{db87a}  Bohm, D., Hidden Variables and the Implicate Order, in {\em Quantum Implications: Essays in Honour of David Bohm}, ed Hiley, B. J. and Peat, D. p.33-45, Routledge \& Kegan Paul, London, 1987.

\bibitem{dbbh91} Bohm, D. and Hiley, B. J., On the Relativistic Invariance of a Quantum Theory Based on Beables, {\em Found. Phys}. {\bf 21}(1991) 243-50.

\bibitem{bh93} Bohm, D. and Hiley, B. J.,  {\em The Undivided Universe: an Ontological Interpretation of Quantum Theory}, Routledge, London  1993.

\bibitem{bh93a} Bohm, D. and Hiley, B. J.,  {\em The Undivided Universe: an Ontological Interpretation of Quantum Theory}, p. 5, Routledge, London  1993.


\bibitem{nb61a} Bohr, N.,  {\em Atomic Physics and Human Knowledge}, p. 61, Science Editions, New York, 1961.

\bibitem{nb61b} ibid, p. 73.

\bibitem{mbpj25}  Born M., Jordan P., Zur Quantenmechanik, {\em Z. Phys}. {\bf 34}, (1925)  858-888.

\bibitem{mb04}  Brown, M., {\em The symplectic and metaplectic groups in quantum mechanics and the Bohm interpretation}, PhD Thesis, London University, 2004. Copy available from bbk.ac.uk/tpru/MelvinBrown

\bibitem{mbbh00}    Brown, M. R. and Hiley, B. J., 2000 Schršdinger revisited: an algebraic approach. quant-ph/0005026.

\bibitem{cz83} Carruthers, P. and Zachariasen, F., Quantum collision theory with phase-space distributions, {\em Rev. Mod. Phys.}, {\bf 55} (1983)  245-85.

\bibitem{wc82}  Clifford W.K., {\em Mathematical Papers}, XLII, Further note on biquaternions, 385-94, Macmillan, London, 1882. 



\bibitem{ac90}  Crumeyrolle A.,    {\em Orthogonal and Symplectic Clifford Algebras:  Spinor Structures}, Kluwer, Dordrecht, 1990.

\bibitem{jd97}  Demaret, J., Heller, M. and Lambert, D., Local and Global Properties of the world, {\em Foundations of Science}, {\bf 2}, (1997), 137-176. 


\bibitem{cdbh}  Dewdney, C., and Hiley, B. J.,  A Quantum Potential Description of One-dimensional Time-dependent  Scattering from Square Barriers,  {\em Found. Phys.}, {\bf 12}, (1982), 27-48.

\bibitem{cdjv88} Dewdney, C., Holland, P. R., Kyprianidis, A. and Vigier, J-P., Spin and non-locality in quantum mechanics, {\em Nature}, {\bf 336} (1988) 536-44.

\bibitem{cdal93} Doran, C. and Lasenby, A., States and Operators in Spacetime Algebra, {\em Found. Phys.}, {\bf 23} (1993) 1295-1327.

\bibitem{ge72} Emch, G. G., {\it Algebraic Methods in Statistical Mechanics and Quantum Field Theory}, Wiley-Interscience, New York, 1972.


\bibitem{if52}  F\'{e}nyes, I. Eine wahrscheinlichkeitstheoretische BegrŸndung und Interpretation der quantenmechanik, {\em  Zeits. f\"{u}r Physik A Hadrons and Nuclei} {\bf 132} (1952) 81-106.

\bibitem{rf87} Feynman, R. P., Negative Probability, in {\em Quantum Implications: Essays in Honour of David Bohm}, ed Hiley, B. J. and Peat, F. D., pp.235-48, Routledge and Kegan Paul, London, 1987

\bibitem{rfbh13} Flack, R., Hiley, B. J., Clarke, J. and Marinov, K., Proposal: To demonstrate the process of weak measurement for atoms using a modified Stern-Garlach apparatus, preprint 2013.


\bibitem{ffbh80} Frescura, F. A. M. and Hiley, B. J., The Implicate Order, Algebras, and the Spinor, {\em Found. Phys.}, {\bf 10}, (1980), 7-31.

\bibitem{ffbh80a} Frescura, F. A. M. and Hiley, B. J., (1980) The Algebraization of Quantum Mechanics and the Implicate Order, {\em Found. Phys.}, {\bf 10}, (1980), 705-22.

\bibitem{ffbh84} Frescura, F. A. M. and Hiley, B. J.,  Algebras, Quantum Theory and Pre-Space, {\em Revista Brasilera de Fisica, Volume Especial, Os 70 anos de Mario Schonberg}, (1984), 49-86.

\bibitem{mg01} de Gosson, M., { \em The Principles of Newtonian and Quantum Mechanics: The Need for Planck's Constant}, Imperial College Press, London, 2001.

\bibitem{mdg10} de Gosson, M., {\em Symplectic Methods in Harmonic analysis and in Mathematical Physics,} Birkh\"{a}user Verlag, Basel, 2010.

\bibitem{mgbh11}  de Gosson, M., and Hiley, B. J., Imprints of the Quantum World in Classical Mechanics, {\em Found. Phys.} {\bf 41}, (2011), 1415-1436. DOI 10.1007/s 10701-011-9544-5, arXiv:1001.4632 v2 [quant-ph]


\bibitem{mg10} de Gosson, M., Quantum Blobs, {\em Found. of Phys.} 
DOI 10.1007/s10701-012-9636-x

\bibitem{hg94} Grassmann, H. G., {\em Gesammeth Math. und Phyk. Werke}, Leipzig 1894.  English translation by Kannenberg, L. C.,  {\em A New Branch of Mathematics: the Ausdehnungslehre of 1844 and Other Works}, Open Court, Chicago, 1995.

\bibitem{sgma07}  Gr\"{o}blacher, S., Paterek, T., Kaltenbaek, R., Brukner, C., Zukowski, M., Aspelmeyer, M. and Zeilinger, A.,  An experimental test of non-local realism, {\em Nature} 446.7138 (2007): 871-875.

\bibitem{vgss}  Guillemin, V. and Sternberg, S.,  {\em Symplectic Techniques in Physics}, Cambridge University Press, Cambridge, 1984.

\bibitem{wh67} Hamilton, W.R., {\em Mathematical Papers}, Vol. 3 Algebra, Cambridge (1967)

\bibitem{rh92} Haag, R.,  {\em Local Quantum Physics}, Springer, Berlin, 1992.


\bibitem{wh58} Heisenberg, W., {\em Physics and Philosophy: the revolution in modern science}, George Allen and Unwin , London, 1958.

\bibitem{dh03} Hestenes, D., Spacetime Physics with Geometric Algebra, {\em Am. J. Phys}, {\bf 71}, (2003), 691-704, 

\bibitem{dhrg71} Hestenes, D., and Gurtler, R.,  Local Observables in Quantum Theory, {\em Am. J. Phys}. {\bf 39}, (1971) 1028-38.


\bibitem{bh04} Hiley, B. J.,Phase Space Description of Quantum Phenomena, in {\em Quantum Theory : Reconsiderations of Foundations-2,} ed. Krennikov, A., pp. 267-86, V\"{a}xj\"{o} University Press, V\"{a}xj\"{o}, Sweden, 2004.

\bibitem{bhbc12} Hiley, B. J., and Callaghan, R. E.,  Clifford Algebras and the Dirac-Bohm Quantum Hamilton-Jacobi Equation. {\em Foundations of Physics}, {\bf 42} (2012) 192-208.
 DOI:  10.1007/s10701-011-9558-z

\bibitem{bhbc10}  Hiley, B. J., and Callaghan, R. E., The Clifford Algebra approach to Quantum Mechanics A: The Schr\"{o}dinger and Pauli Particles. {\em arXiv:Maths-ph}: 1011.4031.

\bibitem{bhbc10a} Hiley, B. J., and Callaghan, R. E., The Clifford Algebra Approach to Quantum Mechanics B: The Dirac Particle and its relation to the Bohm Approach. {\em arXiv: Maths-ph}:1011.4033

\bibitem{bh10b} Hiley, B. J., On the Relationship between the Wigner-Moyal and Bohm Approaches to Quantum Mechanics: A step to a more General Theory?   {\em Found. Phys.} {\bf 40}, (2010) 356-367.  DOI 10.1007/s10701-009-9320-y

\bibitem{bh11} Hiley, B. J., Process, Distinction, Groupoids and Clifford Algebras: an Alternative View of the Quantum Formalism, in {\em New Structures for Physics}, ed Coecke, B., Lecture Notes in Physics, vol. 813, pp. 705-750, Springer (2011).

\bibitem{bh12}  Hiley, B. J., Weak Values:Approach through the Clifford and Moyal Algebras, {\em J. Phys.: Conference Series}, {\bf 361} (2012) 012014. Doi: 10.1088/1742-6596/361/1/012014.  quant-ph/1111.6536.

\bibitem{bh13} Hiley, B. J., On the Relationship between the Moyal Algebra and the Quantum Operator Algebra of von Neumann, arXiv 1211.2098.

\bibitem{jhacwp}  Hirschfelder, J. O., Christoph, A. C. and Palke, W. E., Quantum mechanical streamlines, I Square potential barrier, {\em J. Chem. Phys.}, {\bf 61} (1974) 5435-55.

\bibitem{jhcglb} Hirschfelder, J. O., Goebel, C. J. and Bruch, L. W., Quantized vortices aroubd wavefunction nodes, II, {\em J. Chem. Phys.}, {\bf 61} (1974) 5456-59.

\bibitem{jh78} Hirschfelder, J. O., Quantum Mechanical Equations of Change. I, {\em J.Chem. Phys.}, {\bf 68}, (1978) 5151-62.

\bibitem{ahph01}  Hirshfeld, A. C. and Henselder, P., Deformation quantisation in the teaching of quantum mechanics, {\em Am. J. Phys.}, {\bf 70} (2002) 537-547.

\bibitem{skms11} Kocis, S., Braverman, B., Ravets, S., Stevens, M. J., Mirin, R. P., Shalm, L.K., Steinberg, M. A., Observing the Average Trajectories of Single Photons in a Two-Slit Interferometer, {\em Science}. {bf 332} (2011) 1170-73.


\bibitem{al03} Leggett, A. J., Nonlocal hidden-variable theories and quantum mechanics: An incompatibility theorem,  {\em Found. of Phys}. {\bf 33} (2003) 1469-1493.

\bibitem{vj03} Jones, V. F. (2003), {\em von Neumann algebras}. Lecture Notes from http://www. math. berkeley. edu/~ vfr/MATH20909/VonNeumann2009. pdf.


\bibitem{rl05} Leavens, C. R., Weak Measurements from the point of view of Bohmian Mechanics, {\em Found. Phys}., {\bf 35} (2005) 469-91 doi: 10.1007/s10701-004-1984-8

\bibitem{jm49} Moyal, J. E., Quantum Mechanics as a Statistical Theory, {\em Proc. Camb. Phil. Soc}. {\bf 45}, (1949), 99-123.

\bibitem{wp79} Pauli, W., {\em Wissenschaftlicher Briefwechsel mit Bohr, Einstein, Heisenberg}, Band I, 1919-1929, ed. Hermann, A., Meyenn, K. V. and Weisskopf, V. F., p. 347, Springer, Berlin, 1979.


\bibitem{rp71} Penrose, R., Angular Momentum: a combinatorial approach to Space-time, in {\em Quantum Theory and Beyond}, ed. Bastin, T., Cambridge Uni. Press, Cambridge, 151- 180, 1971.

\bibitem{rp67} Penrose, R.,  Twistor Algebra,  {\em J. Maths Phys}., {\bf 8}, (1967), 345-366.  

\bibitem{pdh}  Philippidis, C., Dewdney, D., and Hiley, B. J.,  Quantum Interference and the Quantum Potential,   {\em Il Nuovo Cimento},  {\bf 52B}, (1979), 15-28.

\bibitem{mr96}  Redei, M., Why John von Neumann did not like the Hilbert Space Formalism of Quantum Mechanics (and What he Liked Instead.) {\em Stud. His. Phil. Mod. Phys}., {\bf 27B}, 493-510, 1996


\bibitem{vn55} von Neumann, J.  {\em Mathematical Foundations of Quantum Mechanics},  Princeton University Press, Princeton, 1955.

\bibitem{vn31}  von Neumann, J., Die Eindeutigkeit der Schr\"{o}dingerschen Operatoren, {\em Math. Ann.}, {\bf 104} (1931) 570-87.

\bibitem{vn35} von Neumann, J., Letter to G. Birkhoff,. November 3, 1935, unpublished.  Cited in Birkhoff, G., Lattices in {\em Applied Mathematics, Symposium on Lattice Theory}, Am. Math. Soc. {\bf XXV}, 158, (1961).]

\bibitem{tt52} Takabayasi, T., On the formulation of quantum mechanics associated with classical pictures, {\em Prog. Theor. Phys.}, {\bf 8} (1952) 143-82: and Remarks on the formulation of quantum mechanics with classical pictures and on relations between linear scalar fields and hydrodynamical fields, {\em Prog. Theor. Phys.}, {\bf 9} (1953) 187.

\bibitem{jv53} Vigier's suggestions are discussed in de Broglie, L., {\em La Physique Quantique Restera-t-elle Indeterministe}, Gauthier-Villiars. Paris, 1953.

\bibitem{ww53}  Weizel, W., Ableitung der Quantentheorie aus einem klassischen, kausal determinierten Modell. {\em Zeits. Physik A Hadrons and Nuclei} {\bf 134} (1953): 264-285: and  Ableitung der Quantentheorie aus einem klassischen Modell. II,  {\em Zeits. Physik A Hadrons and Nuclei} {\bf 135}, (1953): 270-273.



\bibitem{hw24}  Weyl, H.,  Was ist Materie? In {\em Gesammelte Abhandlungen,} {\bf 2},  ed, Chandrasekharan, K., pp.
486Ð510. Berlin: Springer. 1968.  (First published 1924)

\bibitem{hw28} H. Weyl,  {\em The Theory of Groups and Quantum Mechanics}, p. 242, Dover, London, 1931.

\bibitem{jw91} Wheeler J. A.,  {\em At Home in the Universe,} AIP Press, New York, 1991.


\bibitem{jldb} Wilson, A., Lowe, J. and Butt, D. K.,  Measurement of the relative planes of polarization of annihilation quanta as a function of separation distance, {\em J. Phys.}, {\bf 2G}, (1975), 613-24.

\bibitem{w05} Wyatt, R. E., {\em Quantum Dynamics with Trajectories: Introduction to Quantum Hydrodynamics}, Springer, 2005.

\bibitem{zfc06}   Zachos, C. K., Fairlie, D. B. and Cuthright, T. L., Overview of Phase-Space Quantization in {\em Quantum Mechanics in Phase Space} ed. Zachos, C. K., Fairlie, D. B. and Cuthright, T. L.,  pp. 7-34, World Scientific, 2005


\end{thebibliography}
\end{document}